\documentclass[journal abbreviation, manuscript]{copernicus}

\usepackage{tabularx}
\usepackage{subcaption}
\usepackage{moresize}
\usepackage{cancel}

\nolinenumbers

\begin{document}

\title{Observations of high-frequency spectral peaks from in-situ waves in ice data: evidence for nonlinear waves in ice triad interactions?}

\Author[1][jean.rblt@proton.me]{Jean}{Rabault}
\Author[2]{Joey}{Voermans}
\Author[3]{Takehiko}{Nose}
\Author[4]{Graig}{Sutherland}
\Author[2]{Alexander}{Babanin}
\Author[3]{Takuji}{Waseda}
\Author[3]{Tsubasa}{Kodaira}
\Author[5]{Atle}{Jensen}
\Author[5]{Lars Willas}{Dreyer}
\Author[6,7]{Øyvind}{Breivik}
\Author[6]{Gaute}{Hope}
\Author[1,5]{Malte}{Müller}
\Author[8]{Zhaohui}{Cheng}
\Author[8]{Lichuan}{Wu}
\Author[9]{Aleksey}{Marchenko}
\Author[10]{Brian}{Ward}
\Author[1,5]{Kai H.}{Christensen}
\Author[11]{Petra}{Heil}
\Author[5]{Karsten}{Trulsen}

\affil[1]{Norwegian Meteorological Institute, Oslo, Norway}
\affil[2]{The University of Melbourne, Melbourne, Australia}
\affil[3]{The University of Tokyo, Kashiwa, Japan}
\affil[4]{Environment Climate Change Canada, Dorval, Canada}
\affil[5]{University of Oslo, Oslo, Norway}
\affil[6]{Norwegian Meteorological Institute, Bergen, Norway}
\affil[7]{The University of Bergen, Bergen, Norway}
\affil[8]{Uppsala University, Uppsala, Sweden}
\affil[9]{The University Centre in Svalbard, Svalbard, Spitsbergen, Norway}
\affil[10]{University of Galway, Galway, Ireland}
\affil[11]{Australian Antarctic Division and Australian Antarctic Program Partnership, Hobart, Australia}




\runningtitle{Triad harmonics observations in waves in ice}

\runningauthor{J. Rabault et al.}

\received{}
\pubdiscuss{} 
\revised{}
\accepted{}
\published{}


\firstpage{1}

\maketitle


\begin{abstract}
The propagation of waves through the marginal ice zone (MIZ) and deeper into pack ice is a key phenomenon that influences the breakup and drift of sea ice. Waves in ice propagation can be characterized by the associated dispersion relation, which describes both the speed and wavelength of the waves and their attenuation. When waves in ice propagate through a solid, non-cracked, thick enough sea ice cover, significant flexural elastic effects can be present in the dispersion relation. This results in a dispersion relation that opens up for 3-wave interactions, also known as wave triads.

Here, we report the observation of high-frequency spectral peaks in the power spectral density of waves in ice spectra. We show, in two timeseries datasets, that the presence of these high-frequency peaks is accompanied by high values for the spectral bicoherence. This is a signature that the high-frequency peak is phase-locked with frequency components in the main spectral energy peak, and a necessary condition for nonlinear coupling to take place. Moreover, we show for a timeseries dataset that includes several closely located sensors that the dispersion relation recovered from a cross-spectrum analysis is compatible with the possible existence of wave triads at the same frequencies for which the bicoherence peak is observed. In addition to these observations in timeseries datasets, we show that similar high-frequency peaks are observed from additional, independent datasets of waves in ice power spectral densities transmitted over Iridium from autonomous buoys.

These results suggest that nonlinear energy transfers between wave in ice spectral components are likely to occur in some waves and sea ice conditions. This may enable redistribution of energy from weakly damped low-frequency waves to more strongly attenuated higher-frequency spectral components, which can contribute to energy dissipation in the ice. However, more data are needed to offer definite conclusions about the practical importance of this effect in real-world conditions. We suggest several in-situ measurements, numerical investigations, and laboratory experiments to further investigate these phenomena.
\end{abstract}

\section{Introduction}

The Marginal Ice Zone (MIZ) is the dynamic transition zone between the open ocean and consolidated pack ice, and it is strongly influenced by ocean waves. These waves can, in particular, lead to breakup of sea ice \citep{collins2015situ,voermans2020experimental,Voermans2026}, which can dramatically alter sea ice conditions and greatly influences sea ice drift \citep{sutherland2022estimating,thomson2022wave} and melting \citep{thomson2014swell,roach2019advances}.

The propagation of waves in ice-covered areas is characterized by the corresponding wave dispersion relation, which in the absence of currents is formulated as an equation $D(k, f)=0$, where $k$ is the wave number modulus (in rad/m), $f$ is the wave frequency (in Hz; the wave angular frequency $\omega = 2 \pi f$ in rad.Hz is often used in place of $f$, and vice versa), and $D$ is the equation of the dispersion relation. In addition, parameters such as the water depth $H$ (in m), water volumetric mass density $\rho_w$ (in kg/m$^3$), and the properties of the sea ice (in particular, the sea ice Young's modulus $E$ in Pa, Poisson coefficient $\nu$, compressive stress $P$ in Pa, ice thickness $h$ in m, sea ice volumetric mass density $\rho_i$ in kg/m$^3$), play a role in the dispersion relation. More physical parameters may enter the dispersion relation if the influence of additional physics is taken into account under specific conditions. The wave number is, in general, a vector $\bold{k}(f) = \bold{k}_r(f) + i \bold{k}_i(f)$, where the real part $\bold{k}_r(f)$ characterizes the relation between the wavelength and the wave frequency, from which the phase and the group speed can be derived, while the imaginary part $\bold{k}_i(f)$ characterizes the wave attenuation.

Numerous papers have discussed the dispersion relation for waves in ice; see, e.g. \citet{squire1995ocean,voermans2021wave}. Often, the real (propagation properties) and imaginary (attenuation) parts of the dispersion relation are considered separately from each other, in which case $k=|\bold{k}_r|$ stands for the real part of the wavenumber, and one may write $\bold{k}$ instead of $\bold{k}_r$ as a shorthand notation. For example, \citet{greenhill1886wave} wrote down (though it may have been known even earlier) a now classical form for the real part of the wave dispersion relation in ice under the assumption that the ice acts as a thin elastic sheet where flexural stiffness effects are present:

\begin{equation}
    \omega^2 = \frac{gk + Bk^5 - Qk^3}{\coth(kH)+kM},
\end{equation}

\noindent where $B=\frac{Eh^3}{12 \rho_w (1-\nu^2)}$, $Q=\frac{Ph}{\rho_w}$, $M=\frac{\rho_i h}{\rho_w}$.

In the following, we will consider the case where there is no significant contribution arising from the compressive stress, and the wave frequency is low enough that the mass loading effect is negligible with respect to other terms, so that the dispersion relation reduces to:

\begin{equation}
    \omega^2 = \frac{gk + Bk^5}{\coth(kH)}.
    \label{eqn_reduced_disprel}
\end{equation}

Unlike deep-water gravity waves, for which 3-wave resonances are kinematically forbidden and nonlinear energy transfers occur through 4-wave interactions, the flexural term (i.e. $Bk^5$) changes the dispersion relation enough that 3-wave resonances become possible. Interestingly, this makes it easy for the nonlinear resonant wave triad necessary conditions between 3 wave components $(1, 2, 3)$, i.e. (using the shorthand notation mentioned above and omitting the $_r$ indexes for simplicity):

\begin{equation}
  \begin{cases}
    f_1+f_2 & =f_3 \\
    \bold{k}_1 + \bold{k}_2 & = \bold{k}_3,
  \end{cases}
  \label{eqn_triads}
\end{equation}

\noindent to be fulfilled \citep{phillips1966dynamics,phillips1981wave,madsen1993bound}, as we will discuss later in this manuscript. Although triads and other resonant interactions have, to the best of our knowledge, not been discussed extensively on the basis of field measurements, a few theoretical works can be found that discuss their possible existence \citep{marchenko1999stability,marchenko1999stability2}, and this was discussed in detail more recently by \citet{pierce2024sum,pierce2025features}. In practice, these studies derive, in specific and idealized cases, evolution equations governed by Manley–Rowe relations \citep{manley1956some,rowe1958some} for waves in ice, which have a generally similar form across quadratically-coupled 3-wave interaction phenomena. In particular, \citet{pierce2024sum} showed through both analytical and numerical simulations that triads can develop over a few tens of wavelengths in ideal conditions and extract significant amounts of energy from the main incoming wave. The only reported observations of high-frequency spectral peaks we know of, and that may be similar to ours, are the ones from \citet{johnson2021observing}, though these seem to match a 4th order, rather than 2nd order (as we observe here, see below) high-frequency spectral peak, and it is uncertain what their origin is and whether they may come from the effect of wind per se, or can be the result of some nonlinear wave energy redistribution.

Regarding wave attenuation, the most widely accepted consensus, based on theoretical arguments, laboratory experiments, and empirical measurements in the ice \citep{wadhams1973attenuation,yu2022,squire2020ocean,loken2021wave,zhao2015modeling, zhao2015wave}, is that the amount of wave energy dissipation at each frequency is proportional to the local amount of wave energy. This leads to an exponential attenuation of the waves as they propagate in the sea ice and is consistent with the wavenumber formalism introduced above, resulting in:

\begin{equation}
    a(f, \bold{x}, t) = a(f, 0, t) | e^{i ( \bold{k}(f) \cdot \bold{x} - \omega t)} | = a(f, 0, t) e^{-k_i(f) x_{\mathrm{eff}}},
\end{equation}

\noindent where $t$ indicates time (which disappears under a stationary regime assumption), and $a(f, \bold{x}, t)$ is the amplitude of the wave component with frequency $f$ at a location $\bold{x}$ and time $t$, corresponding to an effective propagation depth $| \bold{k}_i(f) \cdot \bold{x} | = k_i(f) x_{\mathrm{eff}}$ in the sea ice.

Generally, $k_i(f)$ is an increasing function of wave frequency $f$ \citep{meylan2014situ,yu2022}. Currently, there is general agreement in the literature on the existence of a scaling $k_i(f) \propto f^n$, with the value of the exponent $n$ reported to be typically around 4.5 \citep{yu2022}, although there is some variability in the literature depending on the exact choices for the wave in ice damping parameterization and the dataset(s) over which this relation is fitted.

However, there is still a debate in the literature on what physical mechanism, or combination of mechanisms, physically determines the attenuation coefficient $k_i(f)$, with effects such as scattering \citep{bennetts2010three,kohout2008elastic,montiel2016attenuation,zhao2016diffusion,bennetts2024thin}, viscous dissipation \citep{zhao2015modeling, sutherland2019two, rabault2017measurements, marchenko2018wave}, turbulence \citep{voermans2019wave,smith2020pancake}, floe-floe interaction \citep{herman2019wave, loken2022experiments,smith2020pancake}, viscoelasticity in ice \citep{mosig2015comparison, zhang2021theoretical, zhao2018three, marchenko2021laboratory}, and overwash \citep{pitt2022model,zha2023numerical}, regularly discussed. Although several formulations have been proposed and implemented to parameterize waves in ice damping (for example, at least 10 different parameterizations are available in WaveWatch III, \citep{tolman1991third,rogers2013implementation,KGulla}), and progress has been reported recently using fitted parameterizations based on scaling laws to unify datasets that span several orders of magnitude of waves in ice parameters \citep{yu2022,kousal2025introduction,rogers2026ice}, models are still largely imperfect at representing waves in ice and, in particular, the wave in ice attenuation $k_i(f)$ \citep{KGulla,rabault2024situ}. Possibly, different combinations of dissipation mechanisms may play a role in different wave and ice conditions, making the problem complex and difficult to disentangle and solve in a general way \citep{herman2024apparent}. If triads exist for waves in ice, this may add an additional possible dissipation mechanism by stealing energy from the main lower-frequency spectral components that are weakly attenuated and moving it into higher frequency components where energy dissipation is stronger.

The ongoing discussions and uncertainties about the details of the mechanisms causing the wave attenuation and the exact formula to use highlight the need for more ground-truth direct in-situ observations to help constrain and calibrate theoretical models. This has been the focus of the waves in ice community for many decades \citep{wadhams1973attenuation,doble2006wave, thomson2013sea, kohout2016situ}, and our groups have recently contributed significant amounts of such data \citep{rabault2023dataset,rabault2024position,muller2025distributed,rabault2022tracking} using new generations of open source, low-cost instruments based on the OpenMetBuoy (OMB) and closely related line of designs \citep{rabault2016measurements,rabault2017measurements,rabault2020open,marchenko2019wave,loken2021wave,rabault2022openmetbuoy,kodaira2024affordable,kodaira2022development,hope2025sfy,MoreRoomattheTopHowSmallBuoysHelpRevealtheDetailedDynamicsoftheAirSeaInterface}. The availability of larger in-situ datasets allows, in turn, to perform more detailed and representative analysis of real-world ground-truth data, which has contributed to reveal interesting dynamics \citep{dreyer2024direct,rabault2024situ,nose2024observation,nose2023comparison}.

In the present work, we study several diverse wave in ice datasets that we have gathered over the past ten years. Our goal is to understand outstanding characteristics of these datasets, namely, harmonics observed in the records. In the following, we mean strictly by a harmonic `a wave with a frequency that is a multiple of the fundamental frequency': so the first harmonic has frequency $2f$ where $f$ is the frequency of the fundamental mode. However, we do not imply anything a priori about the wave vector $\bold{k}$ at the fundamental versus the harmonic frequency - when looking at nonlinear waves, that may be bound or free, the wave vector for the harmonic mode may be allowed to take different values depending on the nature of the harmonic. In particular, a `harmonic peak' is defined in the following purely based on the existence of an individual blob (i.e. a blurred, spread out peak) of energy in the wave periodogram centered around a frequency $2f$, when an individual blob of energy also exists centered around $f$.

We focus on validating that these harmonics correspond to a true signal, identifying likely mechanisms for the energy transfer into these harmonics, and finding out whether these correspond to bound or free waves. For two of these datasets, we were able to obtain raw timeseries of the ice motion at 10Hz. For the other datasets, only the wave power spectral density (PSD) transmitted over the Iridium satellite communication network is available. We show that in all of these datasets we can, now and then, observe clear high-frequency peaks in the PSD of the wave in ice elevation. In the case when timeseries are available, we demonstrate that these high-frequency peaks are accompanied by peaks in the bicoherence, and that they are compatible with the effect of nonlinear wave energy transfer through wave triads. This potentially opens for a novel nonlinear energy transfer and ultimately dissipation mechanism, as pointed out above.

The organization of the manuscript is as follows. In section 2, we present the datasets used and the methodology applied to analyze these. In section 3, we present the results of our analysis. In section 4, we discuss our findings and their impacts. Finally, we offer some conclusions and suggest ways to extend our present investigation.

\section{Data and methods}

In this section, we present both the data we use and the methodologies applied to process these data.

\subsection{Deployments considered}

The data used in the following come from several deployments in sea ice under a variety of conditions. Two different categories of data were obtained: in two datasets, we were able to recover the instruments, so full timeseries are obtained. However, in most deployments the instruments could not be recovered and only the wave spectra transmitted by the buoys are available. The datasets are described in more detail in the following.

\subsubsection{Timeseries data: Tempelfjorden 2015 (T15)}

The first and most comprehensive dataset comes from the 2015 waves in ice measurements performed on landfast ice in Tempelfjorden, Svalbard, using a MOXA logger \citep{sutherland2016observations,rabault2016measurements}. A total of 5 Vectornav VN100 IMU (Inertial Measurement Unit)-based sensors were deployed. Of these, the data from three locations have proved to work well from a technical point of view and produce useful data without measurement dropouts: these correspond to the IMUs with IDs VN1005, VN1002, and VN1007 (from closest to the open water to deepest into the ice), which we will consider in the following.

The VN100 data acquired during this deployment contain the raw X, Y and Z (that is, in the reference frame of the VN100 IMUs) sea ice motion data. In addition, the data are obtained by point-wise sampling of the internal readings of individual sensors at 10 Hz, while the sensors themselves run internally at 800 Hz. This means that we need to apply Kalman filtering a posteriori to the data \citep{sutherland2017method} (since the sensors frame of reference is used) and that the noise level is slightly higher than in other deployments (since sampling rather than averaging is used). The data and processing are openly available; see the details in Appendix A. Appendix A provides both the relevant links to the manuscripts and dataset discussing T15 (and following) data, and a short summary of the conditions in which the T15 deployment took place.

So this dataset provides time-synchronized measurements from 3 sensors at 10Hz. The first sensor (ID VN1005) is around 100 m from the ice edge. The second sensor (ID VN1002) is around 65 m farther in the ice than ID VN1005 along the wave propagation direction. The third sensor (ID VN1007) is around 7 m farther in the ice than ID VN1002 along the wave propagation direction. This dataset makes it possible to perform both bicoherence and cross-spectrum analyses between sensors, as will be discussed below.

The timeseries considered here continuously cover a period of around 14.5 hours. When analyzing these timeseries, the signals are split into segments of length 2$^{14}$ points to make Fast Fourier Transform (FFT) operations fast. This corresponds to around 27 minutes of data per segment. In the following, we will refer to the segment numbers when performing the analysis.

\subsubsection{Timeseries data: Yermak plateau 2020 (Y20)}

The second dataset comes from the OMB-v2018 deployment performed in 2020 over the Yermak plateau from the hovercraft R/H Sabvabaa \citep{rabault2023dataset}. There, a total of six OMB-v2018s, each containing a VN100 IMU, were deployed on sea ice. Later, one of the buoys, with ID 18954, was found and recovered after being stranded on the northwest coast of Iceland \citep{dreyer2024direct}. Therefore, the raw timeseries of the VN100 motion data recorded by this instrument are available from the on-board SD card. The IMU configuration is improved compared to the T15 dataset. More specifically, these VN100 IMU data contain Kalman-filtered (by the processor on board of the IMU package) 10 Hz averaged (by performing an 80-point averaging, instead of individual sample extraction, which provides lower noise) acceleration data in the Earth frame of reference \citep{rabault2020open,dreyer2024direct}. The data are openly available; see the details in Appendix A.

Timeseries are available each time the instrument wakes up to perform a wave measurement, that is, around every 4 hours. Each timeseries is around 21 minutes long and is referred to as a data segment, similar to the T15 deployment. In the following, we will refer to the segment numbers (the numbering is based on the file numbers on the recovered SD card, see the raw data release) when performing the analysis.

\subsubsection{Spectra over Iridium}

The third dataset includes data from Sofar Spotter \citep{raghukumar2019directional}, OMB-v2018 \citep{rabault2020open}, and OMB-v2021 \citep{rabault2022openmetbuoy} deployments, which collected data in both the Arctic and Antarctic. In these deployments, the buoys could not be recovered, so only the 1-dimensional power spectral density (PSD) transmitted over Iridium satellite communications is available.

More specifically, data from the following deployments are used:

\begin{itemize}
    \item The Davis 2020 (D20) Antarctic deployment \citep{rabault2023dataset}. In this deployment, two OMB-v2018 and two Sofar Spotter were deployed on Antarctic landfast ice along a transect perpendicular to the broken ice edge, close to the Davis Research Station.
    \item The Casey 2022 (C22) Antarctic deployment. In this deployment, the OMB-v2021s were deployed on Antarctic landfast ice, close to the Casey Research Station.
    \item The Yermak 2020 Arctic deployment. This is the same deployment as Y20 discussed above \citep{rabault2023dataset}, but data from other buoys are considered.
    \item The 2022 CIRFA-UIT Arctic deployment. This deployment was performed in the Western Fram Strait, East of Greenland. 19 OMB-v2021s were deployed from the R/V Kronprins Haakon. The instruments were deployed in April and May 2022, and data transmissions took place until December 2022 \citep{rabault2024position}.
    \item The 2022 AWI-UTOKYO Arctic deployment. This deployment was performed with 15 OMB-v2021s. The OMBs were deployed on the sea ice northwest of Svalbard in July 2022 and transmitted until October 2022 \citep{rabault2024position}.
\end{itemize}

The goal of considering so many different deployments is to illustrate that high-frequency peaks in the waves PSD can be reliably observed in a variety of locations (covering both the Arctic and Antarctic), sea ice conditions (drift ice and landfast ice of various thicknesses, age, and properties), and seasons, and are not just a particularly rare curiosity. Moreover, three different types of waves in ice instrument have been used to collect these data, relying on different measurement techniques and different signal processing implementations. In particular, while the Sofar Spotter is a closed source design, the Spotter version used to generate the data in the Davis 2020 deployment relies, to the best of our knowledge, on GPS-based Doppler velocity measurements, while the OMBs rely on IMU-based measurements (and the OMB-v2018 and OMB-v2021 use different actual physical sensors, and different implementations of the signal processing routines). As a consequence, given the diversity of instruments, actual hardware sensors and measurement methods, and signal processing implementations used, consistent observation of specific dynamics across these instruments and deployments is very unlikely to be the result of systematic sensors or processing artifacts.

\subsection{Analysis of raw timeseries}

In this subsection, we describe the methodology used to analyze the raw timeseries data available from the T15 and Y20 deployments.

Note that, since all signals considered here are single real-world realizations of an underlying stochastic process (assumed to be stationary within each data segment considered), the Fast Fourier Transform- (FFT-) based formula that we use are strictly speaking only (stochastic) estimators for the underlying "true" characteristics of these signals, which are formally defined based on the underlying (and hidden) autocorrelation and cross-correlation functions (see classical textbook-like introductions to the topic e.g. \citet{bendat2011random}, or summaries such as \url{https://en.wikipedia.org/wiki/Autocorrelation} and \url{https://en.wikipedia.org/wiki/Wiener%E2%80%93Khinchin_theorem}, both accessed 2025-07). However, for simplicity of writing and similar to commonly used conventions in this field, we refer to the FFT-based estimator formulae as the quantities, rather than the estimators for the quantities, that are considered.

\subsubsection{Kalman processing of the T15 data}

For the T15 data, Kalman processing must be applied to the raw IMU data measured in the IMU reference frame to obtain the ice motion in the Earth (North, East, Down) reference frame. This is done using the same open source Kalman filter that is used on the OMB-v2021 \citep{rabault2022openmetbuoy}. The Kalman filter itself is available at \url{https://github.com/adafruit/Adafruit_AHRS} and \url{https://github.com/gauteh/ahrs-fusion} (both accessed 2025-07), and all code and instructions are available along with the data released with this manuscript; see Appendix A.

\subsubsection{Power Spectral Density calculation}

The wave Power Spectral Density (PSD) is calculated from the downward acceleration component in the Earth reference frame. When computing the PSD, we use the Welch method \citep{welch1967use}, typically with an FFT length of 2048 points, 50\% overlap, and a Hann windowing function. This is a standard method for generating the PSD, and is typically similar to what is used in, e.g., the firmware running on board of the OMB-v2018 and OMB-v2021 \citep{rabault2020open,rabault2022openmetbuoy}, and similar buoy designs \citep{kohout2015device}. We also performed consistency checks slightly modifying these parameters during the exploration phase of this work, which confirmed that the results were not affected.

Confidence intervals on PSDs are computed by assuming that the waves timeseries are individual realizations of Gaussian stochastic processes, stationary over the duration of a single wave segment. This is the common assumption made about waves in the ocean (see, e.g., \citet{https://doi.org/10.1029/2025RG000888}), and results in uncertainties for the FFT-based PSD estimator that follow a Chi-square distribution whose number of effective degrees of freedom depends on the parameters used in the Welch transform \citep{welch1967use}. In the context of this work, we wrote a short self tutorial that summarizes the main elements necessary to compute error bars following this model, see \url{https://github.com/jerabaul29/tutorials/blob/main/uncertainty_gaussian_spectrum/uncertainty_gaussian_spectrum.ipynb} (accessed 2026-09-17). While the assumption of a stochastic Gaussian process is not strictly speaking valid for the nonlinear harmonics (i.e., the harmonic blobs parts of the spectra), the short self tutorial above includes relevant Monte Carlo simulations that show that the deviation in the resulting error bars is, in practice for nonlinear harmonics similar to the ones we observe, moderate (at most 10-20\%, which is effectively small in our context). Hence, we use confidence intervals derived from the standard Chi-square law, though we are aware that these are only approximately correct for the harmonic blob. Note that the error bars associated with PSDs are relative to the height of each bin, so that the errors have constant height in semilogy plot, which we take advantage of to condense information in several of our figures. In this manuscript, we present 95\% confidence intervals (corresponding to about 2 standard deviations in the Gaussian framework) for both this and other estimators; stricter confidence intervals could also be used, and do not alter the conclusions.

High-frequency peaks in the PSD of the wave motion are easier to visualize in the acceleration spectra rather than the elevation spectra. This is because, at the hardware level, the actual physical sensors used in the IMUs measure the acceleration, not the elevation, of the waves. Therefore, converting from the acceleration spectrum to the elevation spectrum requires applying the scaling $\mathrm{PSD}_{\eta} = \omega^{-4} \mathrm{PSD}_{\mathrm{acc}}$, where $\mathrm{PSD}_{\eta}$ is the PSD of the elevation and $\mathrm{PSD}_{\mathrm{acc}}$ is the PSD of the acceleration \citep{tucker2001waves}. Thus, the spectral noise level goes from a constant (horizontal line) in $\mathrm{PSD}_{\mathrm{acc}}$ to a sloped line in $\mathrm{PSD}_{\eta}$, which makes it harder to visually distinguish noise from signal in the elevation spectrum. As a consequence, we consistently show acceleration spectra rather than elevation spectra in the following.

\subsubsection{Cross Spectral Density and dispersion relation calculation}

The cross spectral density (CSD) between two adjacent sensors can be used to recover the frequency-dependent phase shift due to the wave propagation delay between the two sensors and hence the dispersion relation \citep{longuet1963observation,tucker2001waves,marchenko2017field}. This analysis can be performed on the T15 dataset; for other datasets, we do not have data from two or more adjacent sensors.

When performing a CSD analysis between two sensors (here, generally indicated by $1$ and $2$) to obtain the dispersion relation, the following procedure is used:

\begin{itemize}
    \item The cross-spectral coherence is calculated following the formula: $\mathrm{CSC}_{1,2}(f) = \frac{|P_{1,2}(f)|^2}{P_{1,1}(f) P_{2,2}(f)}$, where \newline $P_{i,j}(f) = \mathrm{FFT}_i(f) \mathrm{FFT}_j(f)^*$ is the cross-spectral density between the IMUs number $i$ and $j$, $\mathrm{FFT}$ indicates the Fast Fourier Transform operator, and $^*$ indicates complex conjugate. In practice, this is evaluated using the Welch method, i.e. averaging over sub-segments, similar to what is done for the PSD. A coherence value significantly above 0 indicates that phase locking is observed between the two sensors for the frequency considered, and that it is meaningful to extract a phase shift value from the cross-spectral density $P_{1,2}$. Here, we fix the threshold for the cross-spectral coherence at a value of 0.25. This is somewhat low, but this choice is dictated by the low signal-to-noise ratio in the acceleration data, as visible in the PSDs, so that the signal-to-noise ratio of the CSC is also relatively limited. In the frequency domain around the high-frequency peak, where the signal-to-noise ratio goes down to around 5, we even consider a lower coherence value threshold of 0.15, though we acknowledge that this is pushing the limit and the results should be considered with care and only viewed as an indication. However, as visible in Fig. \ref{fig:t15_disprel}, the results are consistent and correspond well to the theory, giving us a posteriori confidence in these relatively low threshold values.
    \item The phase shift of the cross-spectral density between sensors 1 and 2, $\Delta \phi_{1, 2}(f) = \mathrm{phase}(P_{1,2}(f))$, is calculated at the frequencies for which coherence about the threshold is observed. Since the IMUs used in the T15 dataset are generally aligned with the direction of the fjord, which constrains the direction of wave propagation, we calculate the real wavenumber component as: $k_r(f) = \Delta \phi_{1, 2} (f) / D_{1,2}$, with $D_{1,2}$ the distance that separates the two sensors.
\end{itemize}

Similar to PSDs, CSDs and the associated coherence and phase are usually computed under the assumption that the wave timeseries considered are realizations of an underlying stochastic Gaussian process. The distribution for the FFT-based estimators are significantly more complex than for PSDs, and these are discussed in, e.g., \citet{goodman1957joint,carter1973estimation,carter1987coherence,bendat2000random}. We have written a short self tutorial on this topic, available at \url{https://github.com/jerabaul29/tutorials/blob/main/cross_spectrum_coherence_phase/cross_spectrum_coherence_phase.ipynb} (accessed 2026-09-17), which we use as a basis for the calculation of error bars that we present below.

\subsubsection{Bicoherence calculation}

The (auto-, also known as self-) bicoherence \citep{mendel1991tutorial,he2022rossby} is used to estimate the degree of phase-locking between 3 frequencies $f_1$, $f_2$, and $f_3 = f_1+f_2$ in a given timeseries signal. Here, we use the classical definition of the norm-2 bicoherence:

\begin{equation}
    b^2(f_1, f_2) = \frac{|\sum_n F_n(f_1)F_n(f_2)F_n^*(f_1+f_2)|^2}{\sum_n | F_n(f_1)F_n(f_2) |^2 ~ \sum_n | F_n(f_1+f_2) |^2},
\end{equation}

\noindent where the index $n$ indicates the Welch segment considered, and $F_n$ is the FFT of the n-th segment.

Consider the limit case where there is perfect phase locking between components 1, 2, and 3. In this case, all terms in the numerator have the same phase, and the bicoherence will be equal to 1. In the opposite limit case, where there is no phase locking between components 1, 2, and 3, the numerator will average out to 0 given enough samples, and so will the bicoherence. Therefore, a bicoherence of 0 (or close to 0 for a finite and limited number of samples $n$) indicates that there is no phase locking between the components at $f_1$, $f_2$, and $f_3$. The higher the bicoherence value, the more phase locking, with a perfect phase-locking corresponding to a bicoherence of 1 (in practice, a value of 1 is never obtained, due to noise and uncertainties). Note that, as a consequence, a high value of the bicoherence is a necessary but not sufficient condition to demonstrate nonlinear interaction between the frequencies $f_1$, $f_2$, and $f_3$. The bicoherence contains additional information compared with the PSD per se: it is possible to generate signals with similar PSDs but very different bicoherence, as illustrated, for example, in the exercise available at \url{https://github.com/jerabaul29/tutorials/blob/main/Bicoherence/Bicoherence.ipynb} (accessed 2025-07). However, the bicoherence per se is not enough to, e.g., determine the existence, intensity, and direction of possible nonlinear energy transfers.

When bicoherence is significantly higher than $0$ (according to the distribution followed by the bicoherence estimator, see the next paragraph), one can compute the biphase from the formula:

\begin{equation}
  biphase = Phase(\sum_n F_n(f_1)F_n(f_2)F_n^*(f_1+f_2)).
\end{equation}

The biphase effectively measures the relative phase between the 3 modes, and computing it when the bicoherence is significantly above $0$ ensures that the modes are locked relatively to each other, making sure that computing a locked-in phase is meaningful.

Similar to PSDs and CSDs, the distribution of bicoherence and biphase follow specific distributions under the assumption that the timeseries considered are realizations of stochastic Gaussian processes. As for CSDs, the resulting statistics are complex functions, see e.g. \citet{kim1979digital,elgar1989statistics}, and we have written a self-tutorial at \url{https://github.com/jerabaul29/tutorials/blob/main/Bicoherence/Bicoherence.ipynb} (accessed 2026-09-17) which we follow when computing error bars on bicoherence and biphase in the following.

\subsubsection{Triads diagram calculation}

Wave triads, i.e. 3-wave resonant nonlinear interactions, are a possible way through which nonlinear phenomena can transfer energy between wave components. As discussed in the introduction, the necessary conditions for triads to have the possibility to exist are described by Eqn. (\ref{eqn_triads}). Similarly to the bicoherence, these 2 conditions are necessary, but not sufficient, for significant energy to be present in the harmonic following nonlinear wave triad interaction: for example real world effects, such as viscous damping, may prevent significant energy to be transfered and / or accumulated in the harmonic, even when these relations are fulfilled.

The canonical way to visualize possible triad combinations is to plot a two-dimensional triads diagram. For this, the conditions in which both triad conditions are fulfilled are plotted with the longest wave vector ($\bold{k}_3$) starting at the origin and extending along the x-axis. For each position of $\bold{k}_3$ on the x-axis, the locus of admissible $\bold{k}_1$ is then plotted. Therefore, any triangle with vertices (i) the origin, (ii) a locus of admissible $\bold{k}_1$, and (iii) the corresponding $\bold{k}_3$, represents a wave triad condition. For readers unfamiliar with triad diagrams, we provide a short explanation together with a simplified figure in Appendix B. The existence of conditions that enable triads is visible in the results in Fig. \ref{fig:t15_triads}.

We generate the triads diagram for the T15 case in the following way. The dispersion relation that matches most closely the one obtained from the cross-spectrum analysis is selected. Then, we select a set of values for $\bold{k}_3$. For each of these, the upper right quadrant of the two-dimensional plane is discretized numerically, with each point there corresponding to a candidate $\bold{k}_1$ and an associated candidate $\bold{k}_2 = \bold{k}_3 - \bold{k}_1$. For each of these, we then compute the residual $\mathrm{res} = f_1 + f_2 - f_3$, where $f_{1..3}$ is obtained by numerically solving the dispersion relation $D(k_{1..3}, f_{1..3})$ for the corresponding point in the plane. Finally, the locus of $\mathrm{res} = 0$, if it exists, is plotted as a level line. This indicates the locus of admissible $\bold{k}_1$ for which a triad ($\bold{k}_1$, $\bold{k}_3-\bold{k}_1$, $\bold{k}_3$) is obtained.

\subsection{Analysis of data transmitted over Iridium}

The OMB and Sofar Spotter datasets collected over Iridium satellite communications can transmit only a very limited amount of information, typically the 1-dimensional PSD of the wave motion with the OMB series of buoys, as discussed above. This makes it impossible to perform advanced timeseries analysis such as the bicoherence calculation described above.

Therefore, analyzing Iridium-transmitted buoy data is limited to identifying potential high-frequency peaks that may be associated to nonlinear wave interactions. High-frequency peaks in the PSD transmitted by autonomous buoys should be seen mostly as hints, not as a proof, that nonlinear interactions may take place, since it is in theory possible to have high-frequency peaks that do not correspond to a phase-locked mechanism. However, the existence of high-frequency peaks that are not related to the main peak in the PSD requires an arguably unlikely coincidence to take place, which is why we value documenting the recurring existence of such high-frequency peaks.

Note that for the spectra transmitted by OMBs, we know the details of the onboard processing used since all code is open source. As a consequence, we can compute the effective number of degrees of freedom to use in spectra uncertainty estimates, and these are included in the corresponding figures. However, some of the spectra come from the Sofar Spotter \citep{raghukumar2019directional}, which is closed-source and for which we do not know the details of the onboard processing, meaning that we are not able to calculate and include error bars on Sofar Spotter spectra.

\section{Results}

\subsection{Tempelfjorden 2015 (T15)}

\begin{figure}
    \centering
    \includegraphics[width=0.49\linewidth]{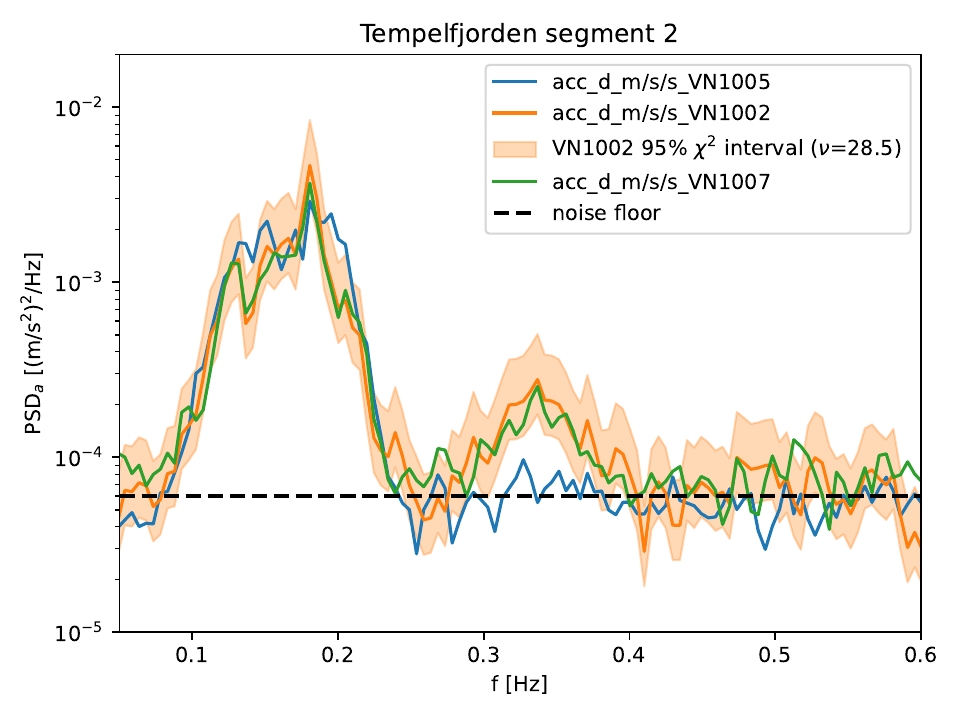}
    \includegraphics[width=0.49\linewidth]{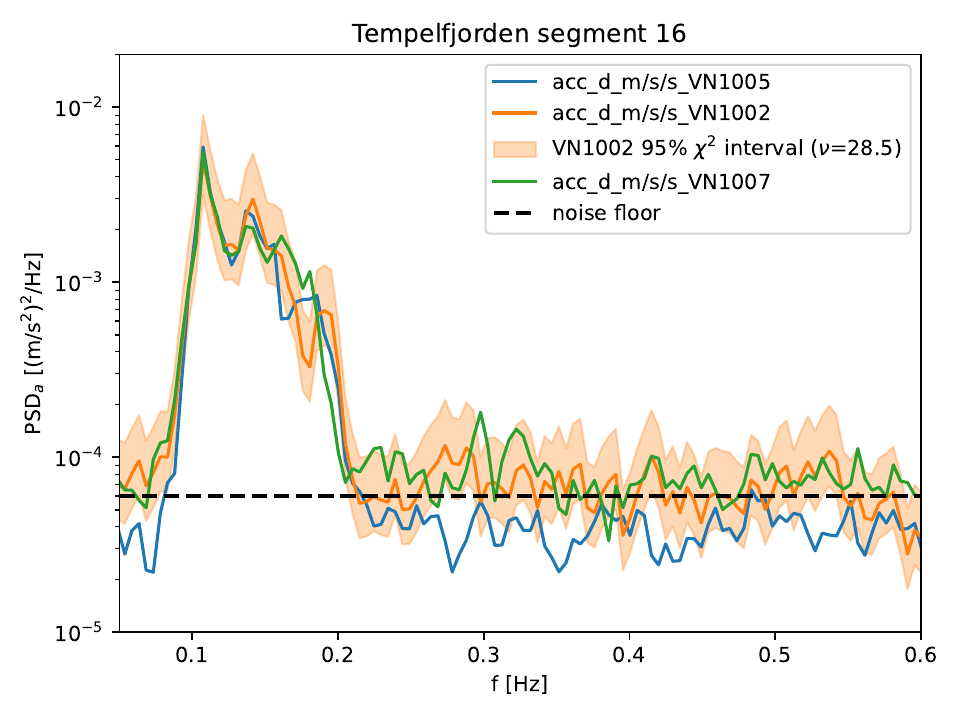}
    \caption{PSDs obtained from the three VN100 IMUs that are part of the T15 dataset. Left: PSDs from the segment number 2. A clear high-frequency peak is observed in IMUs VN1002 and VN1007 (farther in the ice), but not in VN1005 (closest to the open water). The high-frequency peak is relatively broad. Right: PSDs from the segment number 16. No high-frequency peak is observed. Error bars are computed assuming an underlying Gaussian stochastic process (effective number of degrees of freedom $\nu = 28.5$), and shown only for VN1002 to avoid cluttering the figure (the other PSDs are estimated using the same parameters and have similar error bars). The noise floor of the instrument, arising from sensor noise, is clearly visible as a horizontal support line.}
    \label{fig:t15_psd}
\end{figure}

\begin{figure}
    \centering
    \includegraphics[width=0.49\linewidth]{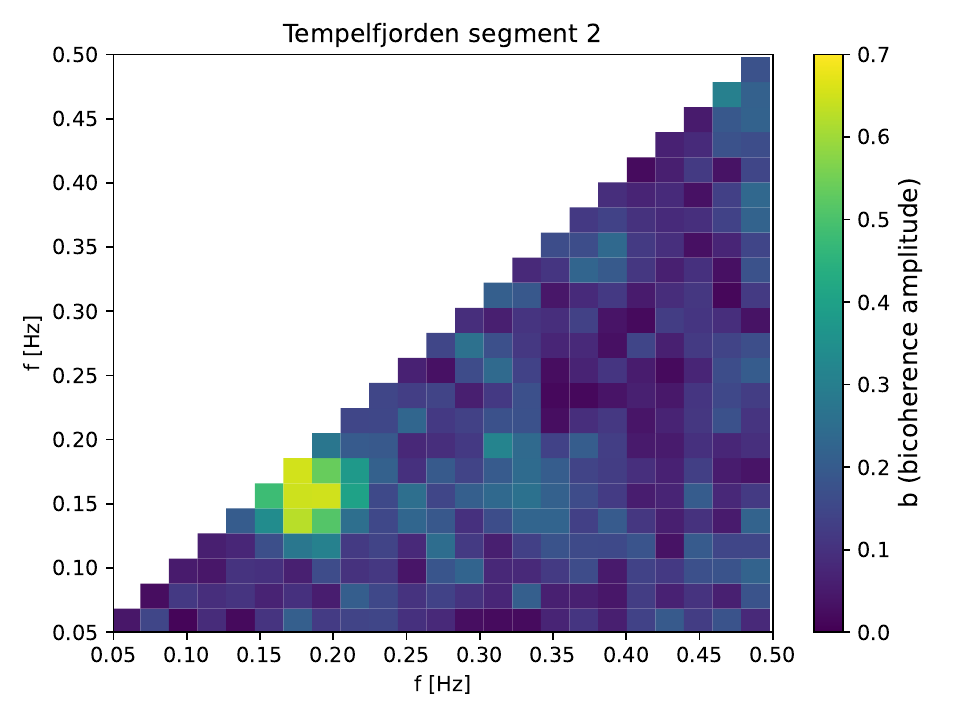}
    \includegraphics[width=0.49\linewidth]{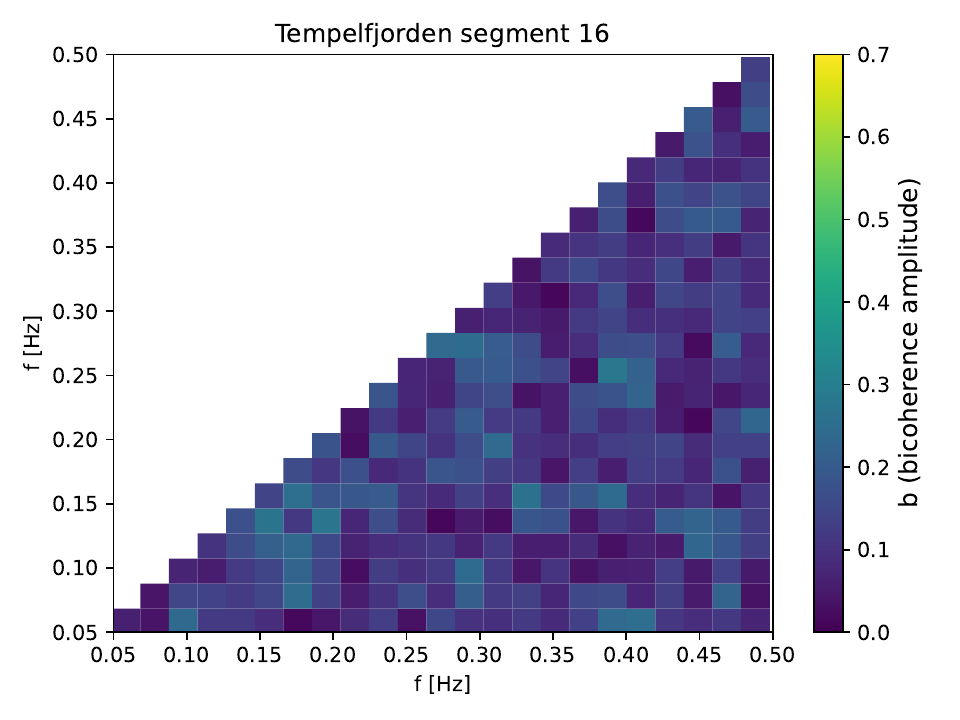}
    \includegraphics[width=0.49\linewidth]{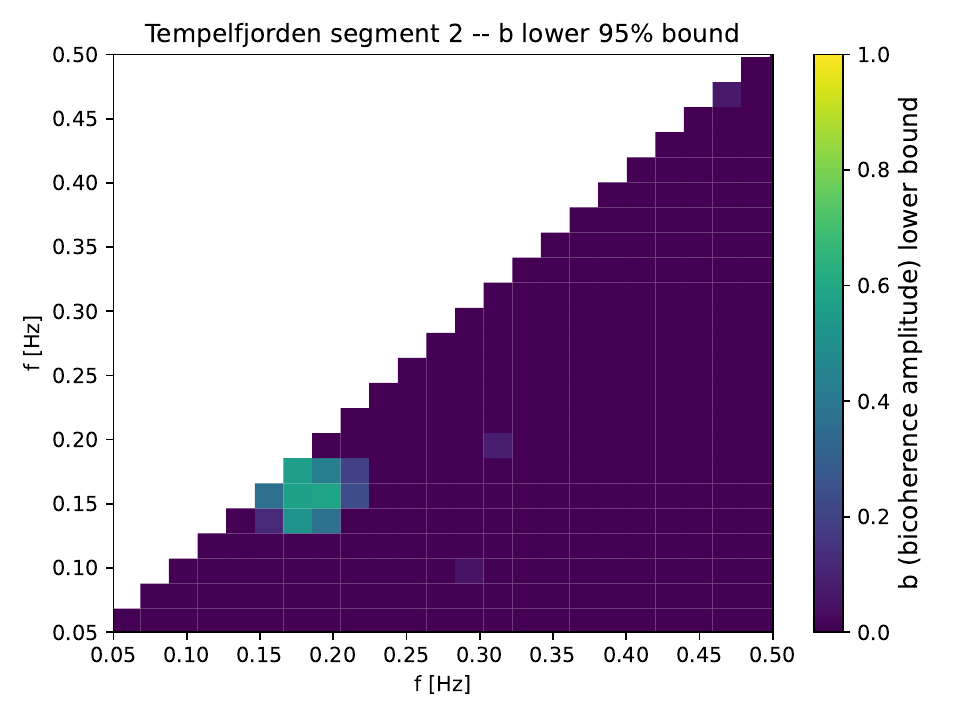}
    \includegraphics[width=0.49\linewidth]{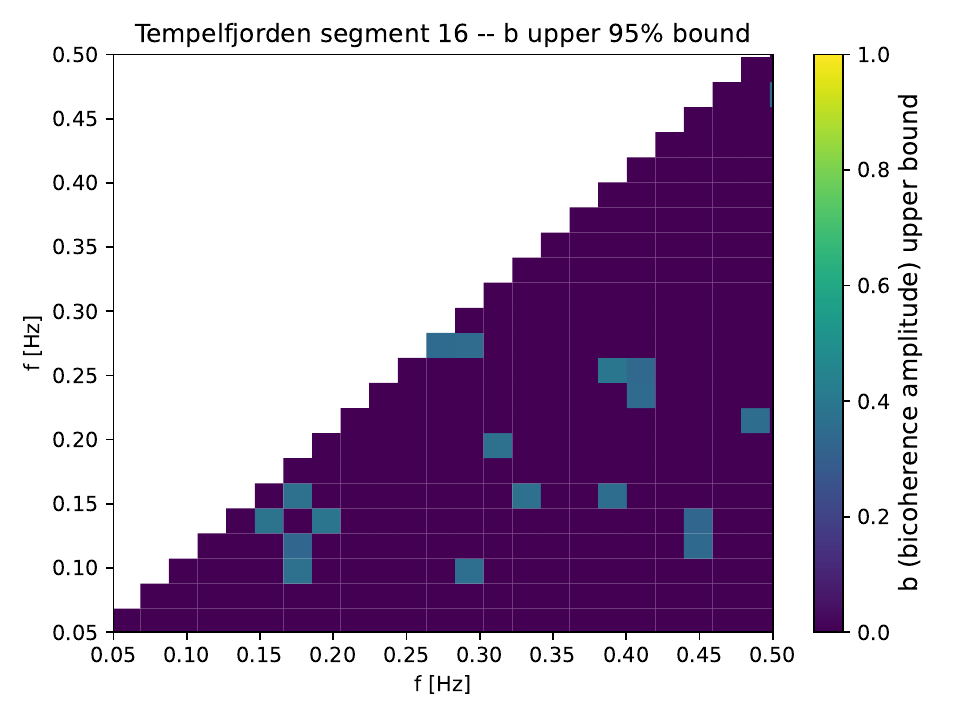}
    \caption{Bicoherence obtained from the IMU VN1002 from the T15 dataset. \\
    \textit{Top row: bicoherence values.} Left: bicoherence from the segment number 2. A clear peak in the bicoherence is observed, typically around the frequencies $f_1 \approx 0.16$, $f_2 \approx 0.18$, resulting in $f_3 = f_1 + f_2 \approx 0.34$, which matches the main and high-frequency peaks in Fig. \ref{fig:t15_psd}. Note that the bicoherence peak is rather a "blob" than a one-point extremum, in good agreement with the observation of a relatively broad high-frequency peak in the PSD plot. Right: bicoherence from the segment number 16. No peak is observed in the bicoherence. \\
    \textit{Bottom row: confidence intervals.} Left: lowest 95\% confidence interval map of the bicoherence matching the plot above. The same bicoherence blob is clearly visible. This confirms that the bicoherence blob observed is well above the level required to be statistically significant. Right: upper 95\% confidence interval map of the bicoherence matching the plot above. Mostly only scattered low-intensity noise-like single peaks are observed, confirming that there is not an as clear statistically meaningful positive bicoherence pattern in this case.}
    \label{fig:t15_bicoherence}
\end{figure}

The T15 dataset contains clear evidence of high-frequency spectral peaks for some of the time segments considered. Fig. \ref{fig:t15_psd} presents an example of a segment with (left, segment 2) and without (right, segment 16) a high-frequency peak. The high-frequency energy blob observed in segment 2 is clearly above the noise floor of the instrument, and also clearly well above the level necessary to ensure that it comes from a physical signal rather than a stochasticity-induced error, as shown by the 95\% confidence regions plotted as a shaded region (note that the y-axis is logarithmic). The corresponding bicoherence plots are presented in Fig. \ref{fig:t15_bicoherence}. This clearly shows that a significant bicoherence peak is obtained in the case where the high-frequency spectral peaks are present. The frequencies associated with the bicoherence peak are typically located around $f_1 \approx 0.16$, $f_2 \approx 0.18$, resulting in $f_3 = f_1 + f_2 \approx 0.34$, which is well matched with the high-frequency peak observed in the PSD. The bicoherence peak is not a sharp 1-point extremum but rather a "blob", which is the expected consequence of considering real world signals that have finite frequency spread and span a localized range of frequencies, rather than a single well defined frequency mode. This indicates that nonlinear interactions may take place over some ranges of frequencies and wavenumber, in good agreement with the observation of a relatively wide high-frequency peak in the PSD. This naturally means that energy tends to spread across neighboring frequencies and form blob-like rather than delta-like signatures. This also means that the term `harmonic' we use in this study is slightly abusive: $f_1$ and $f_2$ are according to these results not equal, and $f_3$ is not strictly speaking double the frequency of an individual single fundamental mode, though it appears to be so in the periodogram due to the finite width of real world spectral energy blobs (hence, the observation of a `harmonic'). However, for simplicity, we use in the following the slightly abusive terminology that $f_3$ is the harmonic of the fundamental frequencies $f_1$ and $f_2$, which are so close in practice for the present case that they are hard to distinguish from a single fundamental frequency. We consider the sensitivity of the triad diagram to variations of these frequencies, among other parameters, later in this manuscript.

Note that the high-frequency peak is observed in the IMUs VN1002 and VN1007, which are farthest into the ice, but not in the IMU VN1005, which is closest to the ice edge. This may indicate either that the high-frequency peak is locally generated in the ice and does not exist close to the open water, or that the high-frequency energy distribution presents a node-and-antinode spatial pattern reminiscent of a standing wave, or that some other complex variation of the spatial distribution of the high-frequency energy content is present in this case.

\begin{figure}
    \centering
    \includegraphics[width=0.49\linewidth]{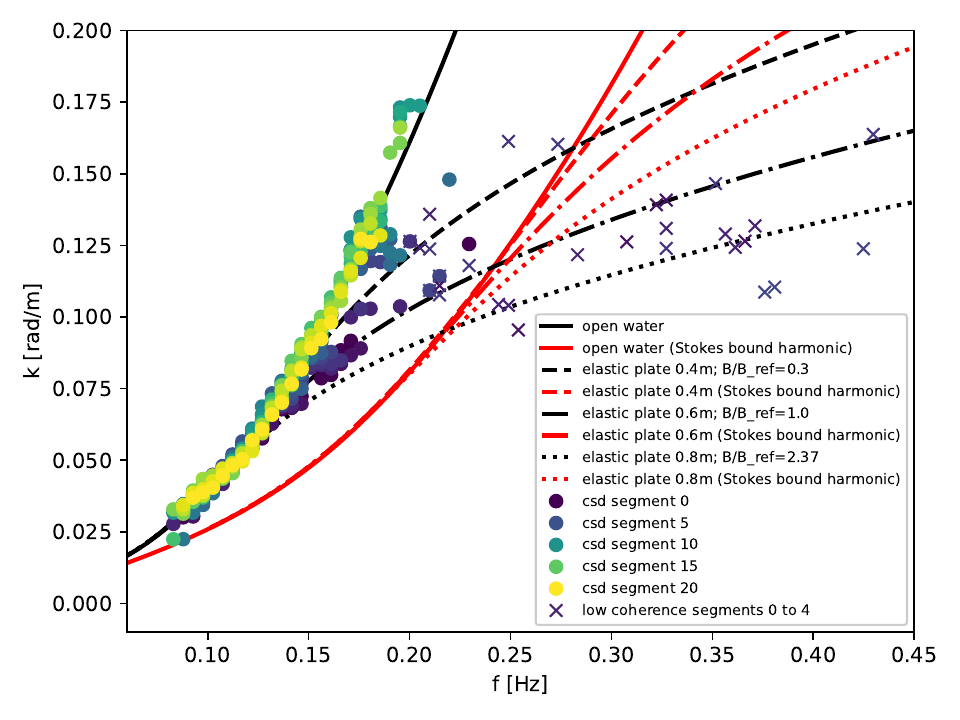}
    \includegraphics[width=0.49\linewidth]{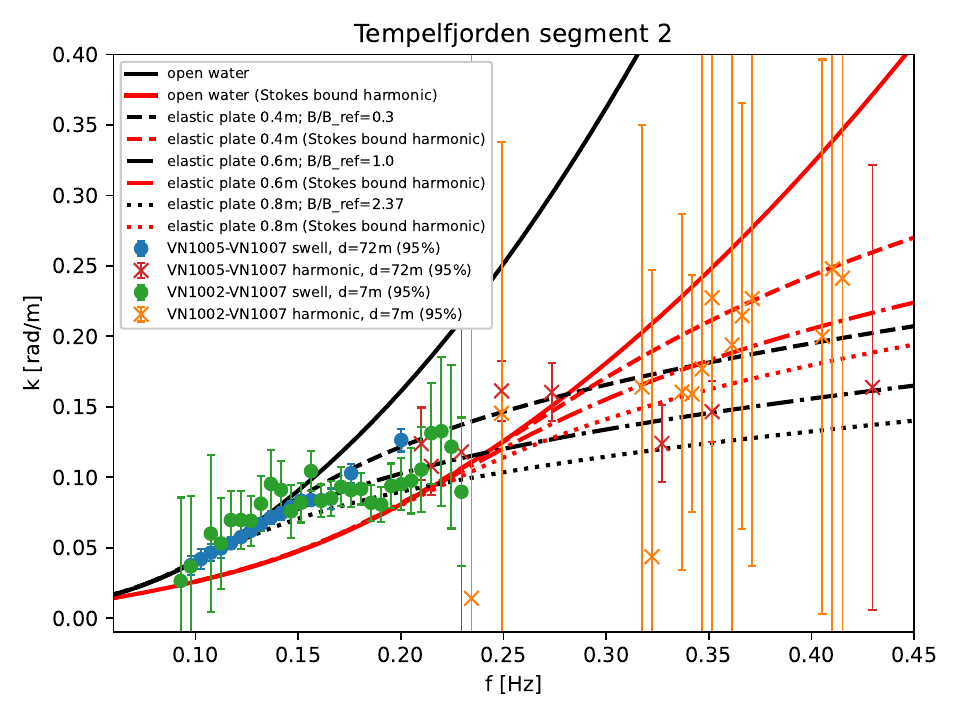}
    \caption{Empirical dispersion relation obtained from a cross-spectral analysis between the IMUs VN1005 and VN1007 (pair with the largest separation distance), and IMUs VN1002 and VN1007 (pair with the shorter separation distance, both deep into the ice). \\
    \textit{Left panel: comparison across all segments for the sensor pair with the largest separation distance.} Clear deviation from the deep water dispersion relation, in good agreement with the thin elastic plate dispersion relation and with previous results \citep{sutherland2016observations}, is observed until around segment 5. For later segments, the empirical dispersion relation matches the deep water one. Here, we show both the empirical dispersion relation obtained for the main PSD peak (large dots), and for the higher frequency components when these are present in the PSD (crosses; due to the low signal-to-noise ratio in the higher frequencies, these are obtained using a low value of the CSC threshold, and are therefore particularly noisy and should be taken only as an indication). The best-matching dispersion relation, which corresponds to an ice thickness of around 0.6 meters with the default ice parameters listed in the text (in good agreement with the field measurements reported in \citet{sutherland2016observations}), provides the $B_{\mathrm{ref}}$ value for the $B$ coefficient in Eqn. (\ref{eqn_reduced_disprel}). The points observed do not seem to match the Stokes bound harmonics. Error bars are not included to not clutter the figure. \\
    \textit{Right panel: analysis on segment 2 for both pairs of sensors, including error bars.} This shows the data for segment 2 alone, including error bars obtained from propagating the uncertainty computed for the cross coherence phase. The error bars on the pair of sensors with the longest separation distance are tight enough to confirm the validity and robustness of the results presented in the left panel. The error bars on the pair of sensors with the shorter separation distance are much larger, and while they are generally compatible with the results from the pair with longest separation distance (minor differences regarding the effective ice thickness to be used are physical, as ice thickness varied on the field between the sensors closest to the open water (around 60cm ice thickness) and the ones further in the ice (closer to 80cm at some places)) and offer a cross validation of the results, they would not be enough, on their own, to draw any conclusions regarding the higher frequency part ($f>0.20$Hz) of the spectrum.}
    \label{fig:t15_disprel}
\end{figure}

In order to further investigate whether the high-frequency spectral peak and bicoherence peak can correspond to a nonlinear wave interaction caused by wave triads, we compute the real-world wave dispersion relation by applying the cross-spectral density analysis between the two sensors furthest apart (VN1005 and VN1007), and the two sensors that are close to each other inside the ice and both display a higher bicoherence blob (VN1002 and VN1007). The two IMUs that are located furthest away from each other give a higher relative accuracy in the phase-shift measurements at the cost of a lower coherence, while the two IMUs closest to each other have a higher coherence but the short separation means that the relative wave phase uncertainty is high. Due to the separation distance, phase aliasing is observed between the two sensors further apart. This phase aliasing is removed by enforcing a smooth monotonic increase in $\Delta \phi$ for increasing frequencies, by applying suitable corrections by $2 \pi$ when aliasing jumps take place. The results are presented in Fig. \ref{fig:t15_disprel}, with the dispersion relation from Eqn. (\ref{eqn_reduced_disprel}) included for different effective sea ice thickness values and otherwise standard landfast ice mechanical characteristics (we use similar values as \citet{rabault2016measurements}, i.e. $g=9.81$ m/s$^2$, $E=2 \times 10^9$ Pa, $\nu=0.3$, $\rho_w = 1025$ kg/m$^3$, $\rho_i=920$ kg/m$^3$). We also include the Stokes bound harmonic \{$k$, $\omega$\} relations (which are obtained, following the Stokes waves solution, by doubling both $k$ and $\omega$ relative to the dispersion relation of free waves). The left panel of Fig. \ref{fig:t15_disprel} shows the dispersion relation observed, across segments as a function of time (color code of the points), between the two instruments furthest apart. The right panel shows the dispersion relation observed for segment 2 between both pairs of sensors, and includes error bars arising from the coherence uncertainty assuming an underlying stochastic Gaussian process.

As visible in Fig. \ref{fig:t15_disprel} (left, based only on the sensor pair with the largest separation distance), and in good agreement with the findings previously reported in \citet{sutherland2016observations}, a significant deviation from the open water dispersion relation, which matches well with a thin-plate elastic dispersion relation with the same typical thickness as measured in the field, is observed early in the data (up to around segment 5). We show the data points obtained about the low signal-to-noise ratio higher frequency components by using the low CSC threshold as thin crosses, and we expect that these are particularly uncertain and should, therefore, only be taken as an indication. Later on, the dispersion relation collapses back to the deep-water dispersion relation, which we interpret to correspond to the development of cracks through the ice cover. Cracks were indeed observed when the instruments were recovered, while they were not present when the instruments were deployed, consistent with this interpretation. The clustering of the points observed matches well the free wave dispersion relations, and does not appear to match Stokes harmonic solutions, though we concede that the low coherence threshold used to produce the points for the higher frequency part of the plot means that these are only hints, and should not be taken alone as a definite evidence of the nature of the harmonic. Error bars are not included to not clutter the figure.

Fig. \ref{fig:t15_disprel} (right) shows specifically results obtained for segment 2, and includes error bars obtained for each of the two pairs of sensors considered. This confirms that the error bars on the sensor pair with the highest separation distance are tight enough to give confidence in the clustering observations reported on the left panel. In addition, this shows that the results obtained from the pair of sensors with shorter separation distance are too uncertain to draw any categorical conclusion on the dispersion relation for the higher frequency part of the spectra (above typically $0.20$Hz) due to the higher relative uncertainty in the phase shift arising from a smaller denominator in the equation to obtain $k$. However, the observations and their error bars are compatible with the results from the pair of sensors with the largest separation, which can be seen as a general consistency check. Note that the ice thickness was varying in space, from around 60 cm around the sensors closest to the ice edge, up to around 80 cm at some locations around the sensors deepest into the ice, which means that the effective dispersion relation observed may be slightly different between the two pairs of sensors considered.

\begin{figure}
    \centering
    \includegraphics[width=0.95\linewidth]{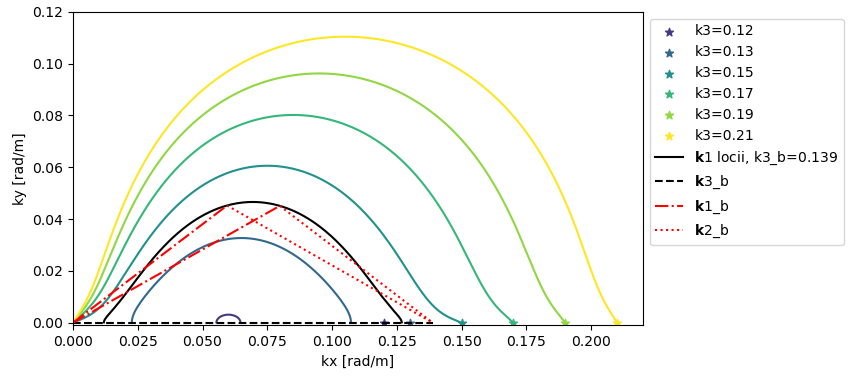}
    \caption{Wave triad diagram corresponding to the thin plate dispersion relation that best matches the empirical dispersion relation observed for early segments in Fig. \ref{fig:t15_disprel} (i.e. elastic plate with thickness 0.6m). This dispersion relation can sustain the wave triad conditions in the frequency range of the high-frequency peak that we observe. For the specific value of $\bold{k}_3$ corresponding to the PSD and bicoherence high-frequency peak, sets of loci for $\bold{k}_1$ and $\bold{k}_2$ that match the main peak in the PSD are identified (red). In this and similar figures, $kx$ and $ky$ indicate the $x$ and $y$ components of a wave vector $\bold{k} = kx \bold{e}_x + ky \bold{e}_y$.}
    \label{fig:t15_triads}
\end{figure}

We use the best-match elastic plate dispersion relation from Fig. \ref{fig:t15_disprel} (i.e., the one for an elastic plate with thickness 0.6m) to perform a wave triad analysis. The corresponding value for the $B$ coefficient in Eqn. (\ref{eqn_reduced_disprel}), is marked as $B_{\mathrm{ref}}$. The results are presented in the wave triad diagram in Fig. \ref{fig:t15_triads}. As visible there, valid loci curves are found meaning that triads are indeed possible given this dispersion relation. The specific locus of $\bold{k}_1$ corresponding to the value of $k_3$ that matches the frequency for the high-frequency peak revealed by the bicoherence analysis is plotted in black. For this locus, we find two sets of wave vectors $(\bold{k}_1, \bold{k}_2)$ that match the frequencies $(f_1, f_2)$ (resp., $(f_2, f_1)$) obtained from the bicoherence peak. These are plotted in red. Therefore, the PSD, bicoherence analysis, dispersion relation derived from the CSD analysis, and wave triads diagram are all consistent with each other and indicate that nonlinear mechanisms corresponding to wave triads are compatible with our observations. Note that all these empirical facts are necessary and not sufficient conditions to establish the existence of wave triads as the mechanism driving the existence of the harmonic blob, as discussed above. Therefore, this is the strongest set of evidence we can provide with the dataset at our disposal, but this is not an indisputable proof that our observations arise due to nonlinear wave triads redistributing energy across spectral components.

Knowing that triads can be expected, one can also observe, back on Fig. \ref{fig:t15_psd}, that there are two significant dips in the main peak of the PSD of the IMUs furthest in (VN1002 and VN1007) relatively to the PSD of the IMU furthest out (VN1005). These dips are observed consistently over several FFT points, for both IMUs further in the ice, at frequencies around $0.16$Hz and $0.18$Hz, which sum up to the typical frequency of the high-frequency peak (around $0.34$Hz). In addition, the depth of these dips is typically similar to or larger than the width of the 95\% confidence intervals, and both dips span several consecutive PSD points (i.e., these are likely not one-off stochastic jitterness). Therefore, these may be a signature of the energy that is being "stolen" from the main peak and transferred into the high-frequency peak by the triad interactions. 

Since these data are obtained from field measurements, there are significant uncertainties on all parameters related to the ice properties (which are known to vary widely in time and space). In addition, the bicoherence peak is a relatively large "blob", and the PSD high-frequency peak is also relatively wide. Therefore, there is no single clear value for these parameters. Moreover, the "peak bicoherence triad" analyzed above corresponds only to the maximum of the bicoherence "blob". In reality, the spectra observed are broad-banded, and the high-frequency peak may rather be the consequence of a continuum of triad interactions than of a single triad. In particular, a full description of the dynamics would require integrating over all possible triads, akin to Hasselmann's equation \citep{hasselmann1962non}. However, this is beyond the scope of the present work, and to offer a convincing comparison between these more advanced descriptions and observations, one may need significantly more detailed field data including measurements from a large array of instruments.

\begin{figure}
    \centering
    \includegraphics[width=0.85\linewidth]{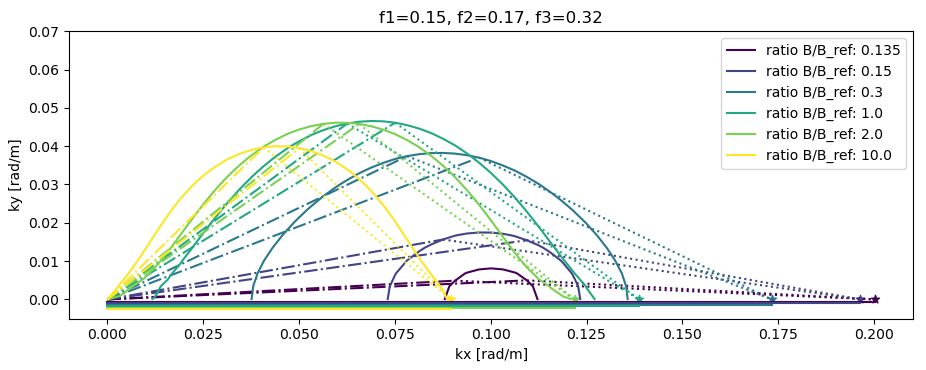}
    \includegraphics[width=0.85\linewidth]{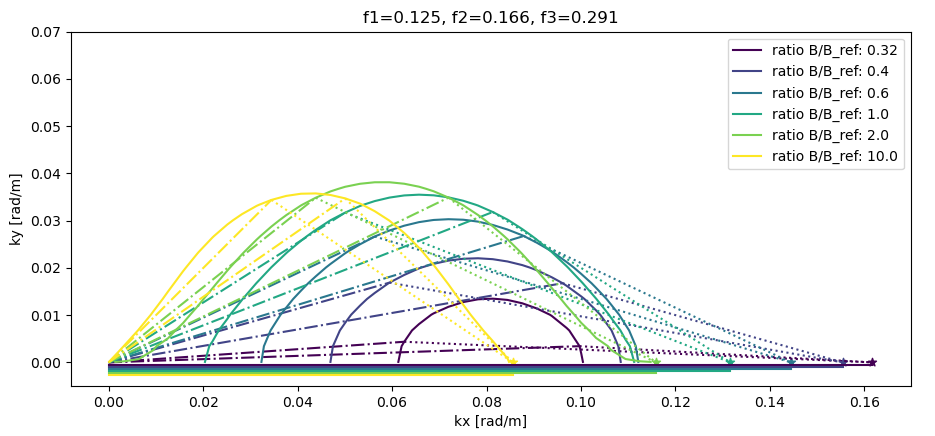}
    \includegraphics[width=0.85\linewidth]{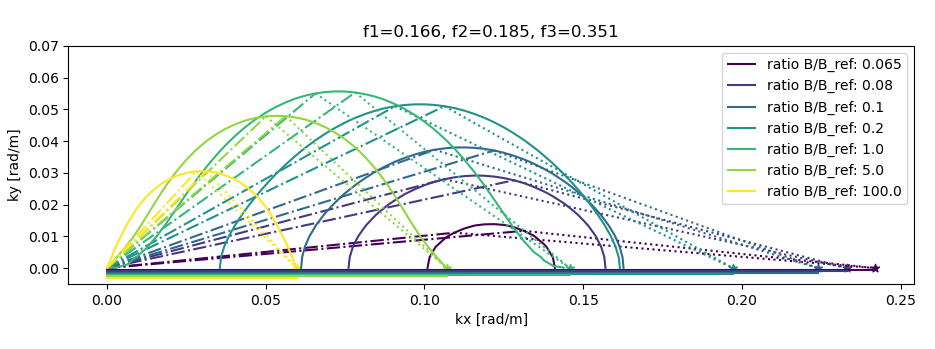}
    \caption{Sensitivity analysis of the triads shapes. The 3 plots indicate the triads found for different values of $f_1$, $f_2$, and $f_3=f_1+f_2$ chosen to span the width of the blob of high bicoherence observed in Fig. \ref{fig:t15_bicoherence}. The series of triad diagrams for each of the plots, coded by the colors, corresponds to different values of $B$ relative to the best fit value $B_{ref}$ found in Fig. \ref{fig:t15_disprel}. Each triad locus curve is for the value of $k_3$ obtained from solving the dispersion relation for $f_3$ using the corresponding value of $B$. This confirms that our results are quite robust to varying all the parameters involved.}
    \label{fig:t15_triads_sensitivity_analysis_Bcoeff}
\end{figure}

In order to double-check the robustness of our analysis, we therefore perform a sensitivity analysis by varying the triad frequencies and the $B$ coefficient in Eqn. (\ref{eqn_reduced_disprel}). The results, which cover the typical range of admissible parameter values, are presented in Fig. \ref{fig:t15_triads_sensitivity_analysis_Bcoeff}. As visible there, our results are robust to variations of the $f_1$, $f_2$, and $f_3=f_1+f_2$ frequencies within the range of values observed within the bicoherence blob, even for quite large changes in the coefficient $B$. This gives us good confidence in the robustness and validity of our results.

\subsection{Yermak plateau 2020 (Y20)}

\begin{figure}
    \centering
    \includegraphics[width=0.48\linewidth]{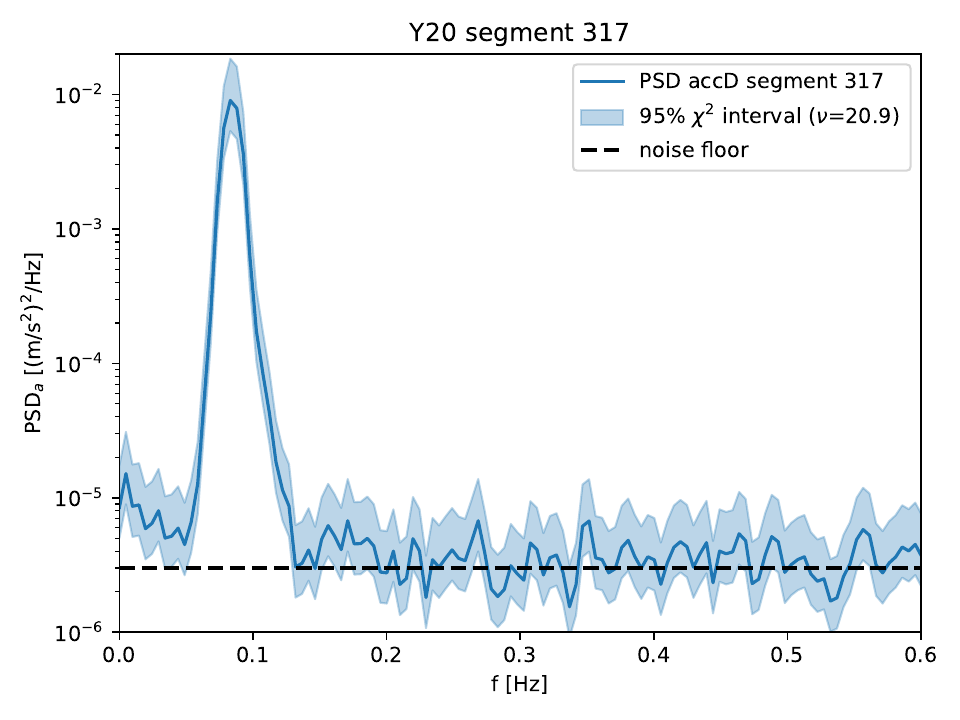}
    \includegraphics[width=0.48\linewidth]{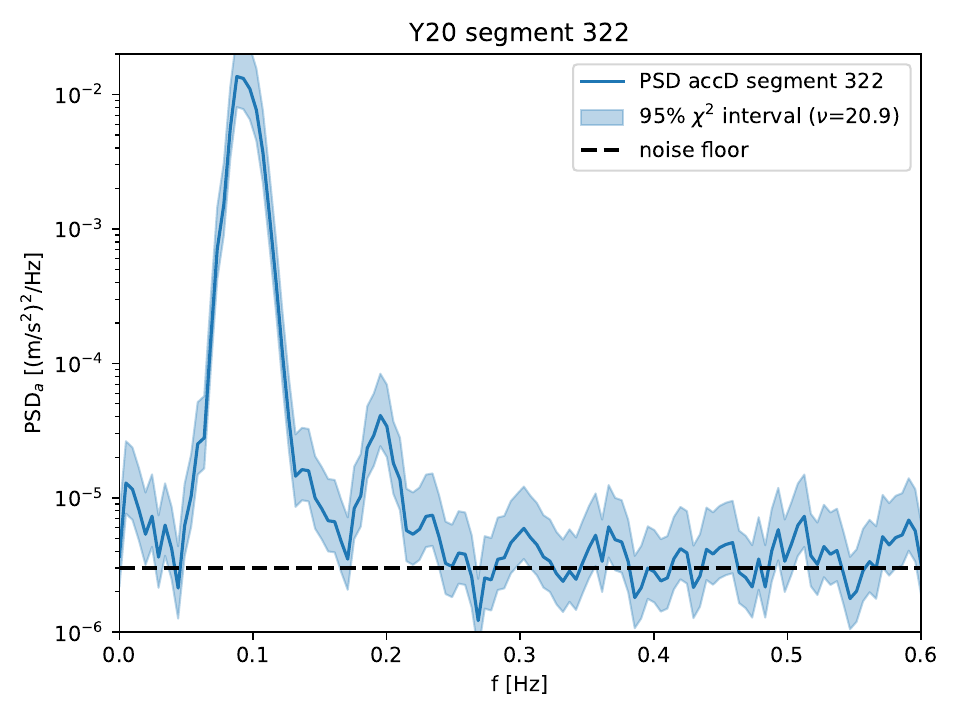}
    \caption{PSD of the vertical acceleration recorded by the VN100 IMU present in the Y20 buoy. Left: example of PSD without high-frequency peak present (corresponding to segment 317). Right: example of PSD with a high-frequency peak (corresponding to segment 322). Error bars are computed assuming an underlying Gaussian stochastic process. The processing chain applied on board the instrument results in shorter timeseries than for T15 data and an effective number of degrees of freedom $\nu = 20.9$, similar but slightly different from the analysis applied on the T15 data segments.}
    \label{fig:y20_psd}
\end{figure}

\begin{figure}
    \centering
    \includegraphics[width=0.48\linewidth]{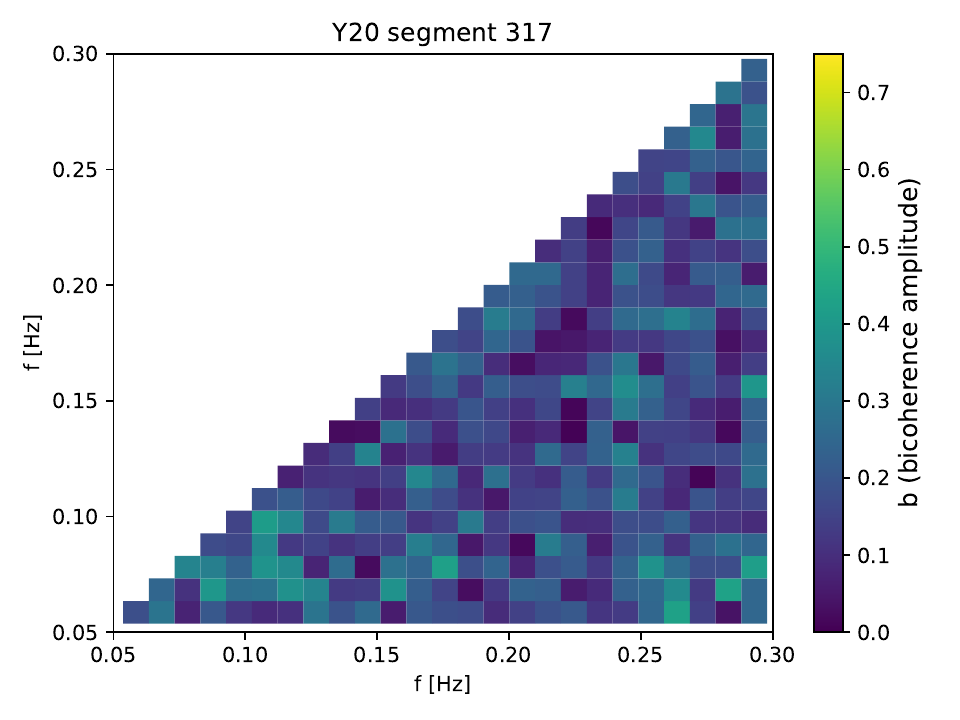}
    \includegraphics[width=0.48\linewidth]{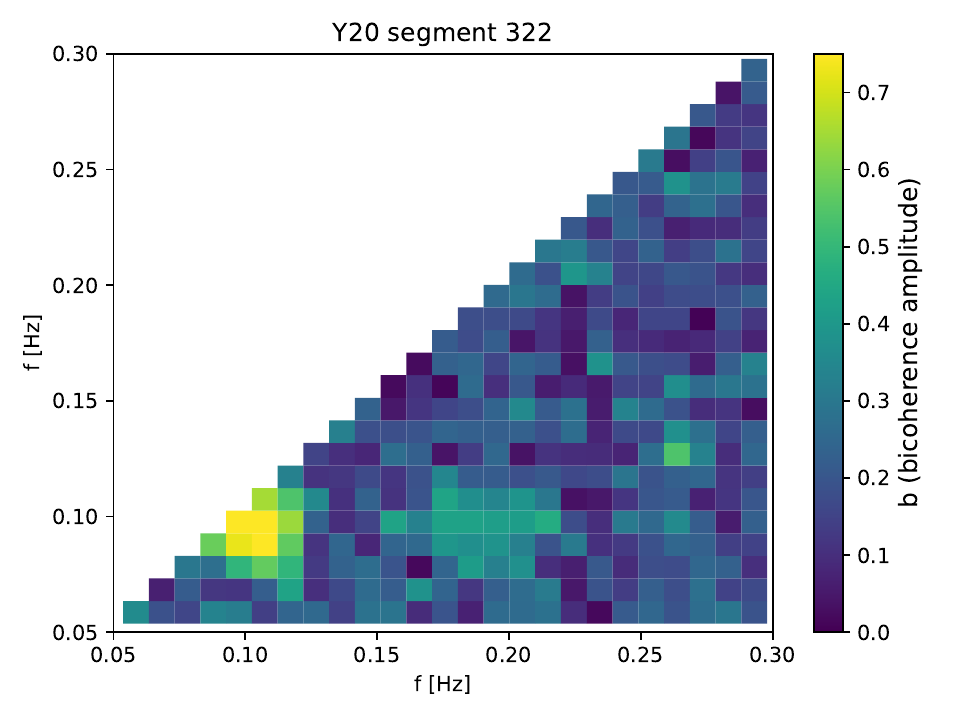}
    \includegraphics[width=0.48\linewidth]{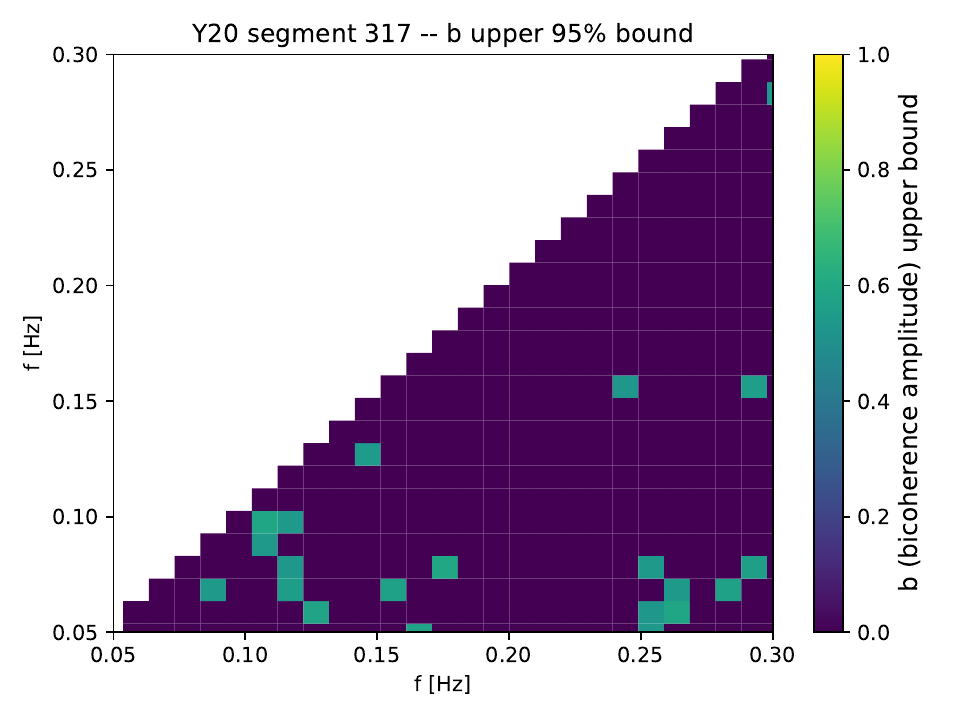}
    \includegraphics[width=0.48\linewidth]{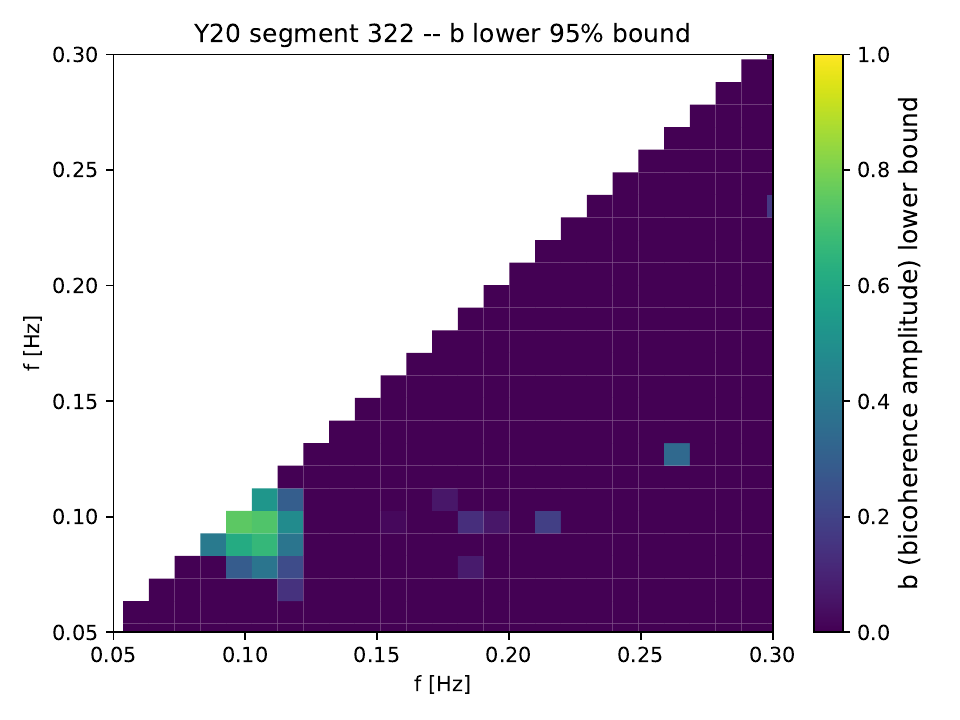}
    \caption{Bicoherence of the vertical acceleration recorded by the VN100 IMU present in the Y20 buoy, presented in a similar way as was done for the T15 data in Fig. \ref{fig:t15_bicoherence}. \\
    \textit{Top row: bicoherence values.} Left: bicoherence for the same segment that does not contain a high-frequency peak as used in Fig. \ref{fig:y20_psd} (segment 317). Right: bicoherence for the same segment that does contain a high-frequency peak as used in Fig. \ref{fig:y20_psd} (segment 322). A clear bicoherence peak, that matches the main PSD peak and the high-frequency peak, are observed for segment 322. \\
    \textit{Bottom row: confidence intervals.} Left: upper 95\% confidence interval map of the bicoherence matching the plot above. Mostly only scattered low-intensity noise-like single peaks are observed, confirming that there is not an as clear statistically meaningful positive bicoherence pattern in this case. Right: lowest 95\% confidence interval map of the bicoherence matching the plot above. The same bicoherence blob is clearly visible. This confirms that the bicoherence blob observed is well above the level required to be statistically significant.
    }
    \label{fig:y20_bicoherence}
\end{figure}

Similarly to the T15 dataset, the Y20 dataset contains clear high-frequency peaks in the PSDs of some data segments, as visible in Fig. \ref{fig:y20_psd}, which compares one case without (left, segment 317) and with (right, segment 322) such a high-frequency peak. The associated bicoherence for the same two segments is presented in Fig. \ref{fig:y20_bicoherence}. A clear bicoherence peak, which corresponds to the high-frequency peak observed in Fig. \ref{fig:y20_psd}, is visible there.

Since only one buoy was deployed, the dispersion relation cannot be recovered from the CSD as was done in the T15 case. Therefore, we have no way to know with certainty how the dispersion relation exactly looks like in this case and what the associated triads diagram is. However, there is no doubt that, similar to the T15 case, a dispersion relation that contains a significant elastic plate contribution will support triads that can be compatible with the bicoherence observed.

\subsection{Spectra over Iridium}

Most available waves in ice data rely on Iridium communications for data collection and report only the 1-dimensional PSD of the wave motion. This makes it impossible to compute the bicoherence. However, it is still possible to identify high-frequency peaks in PSDs, which may indicate candidate events for dynamics similar to what has been identified in the T15 and Y20 datasets.

\begin{figure}
    \centering
    \includegraphics[width=0.33\linewidth]{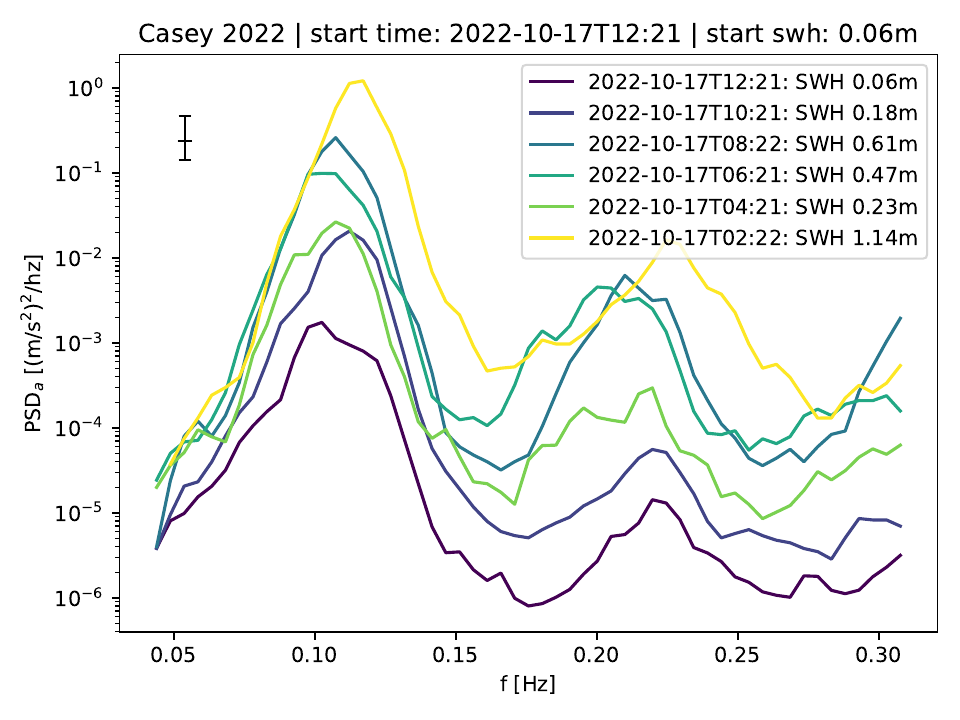}
    \includegraphics[width=0.33\linewidth]{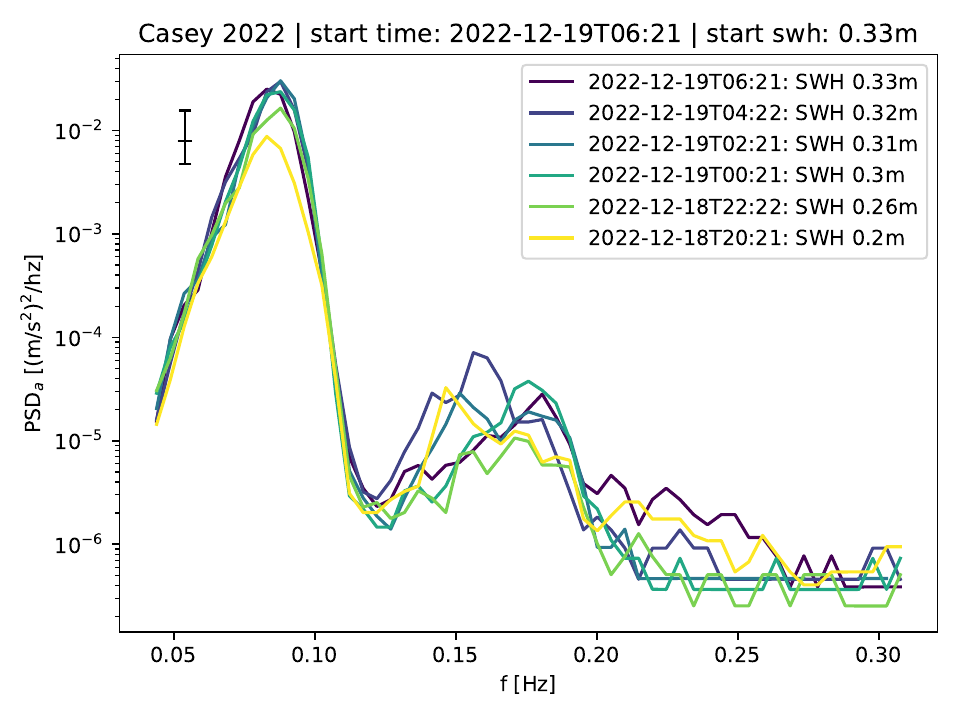}
    \includegraphics[width=0.33\linewidth]{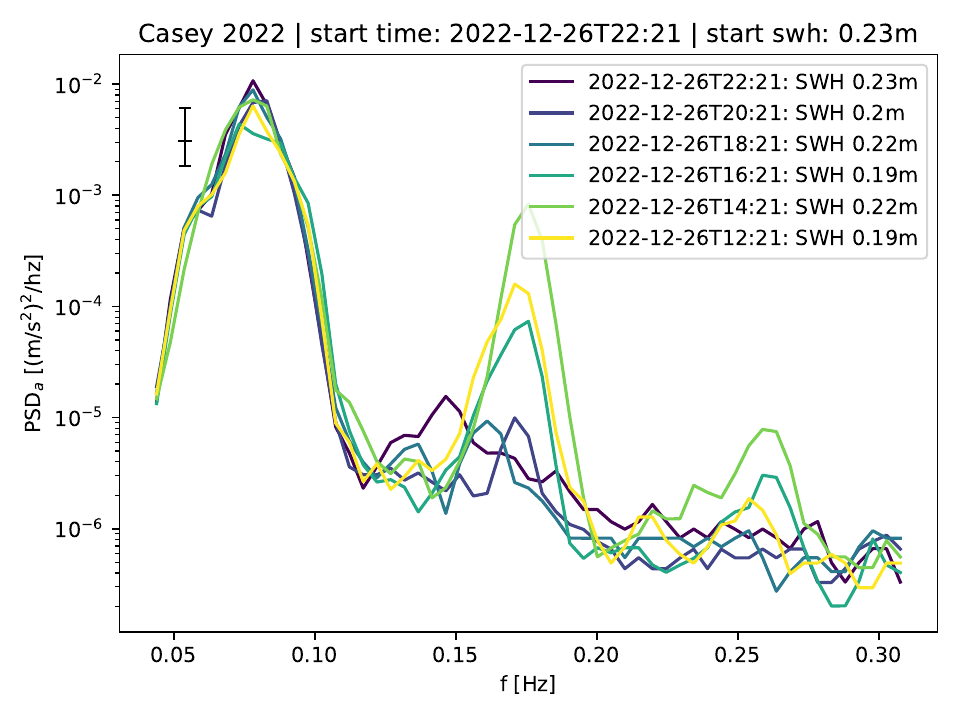}
    \includegraphics[width=0.33\linewidth]{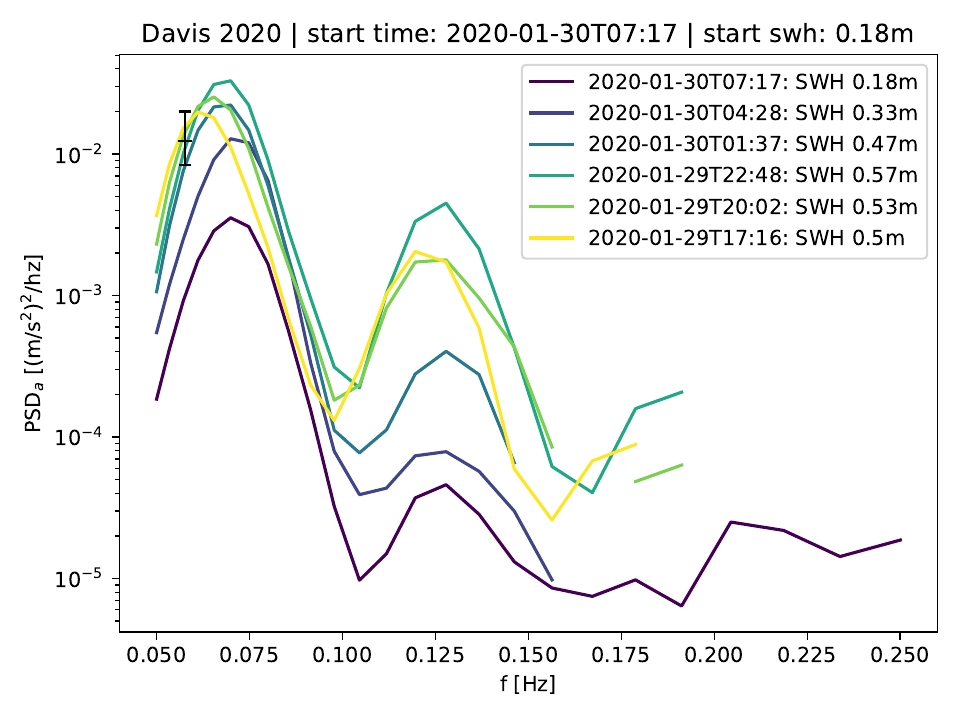}
    \includegraphics[width=0.33\linewidth]{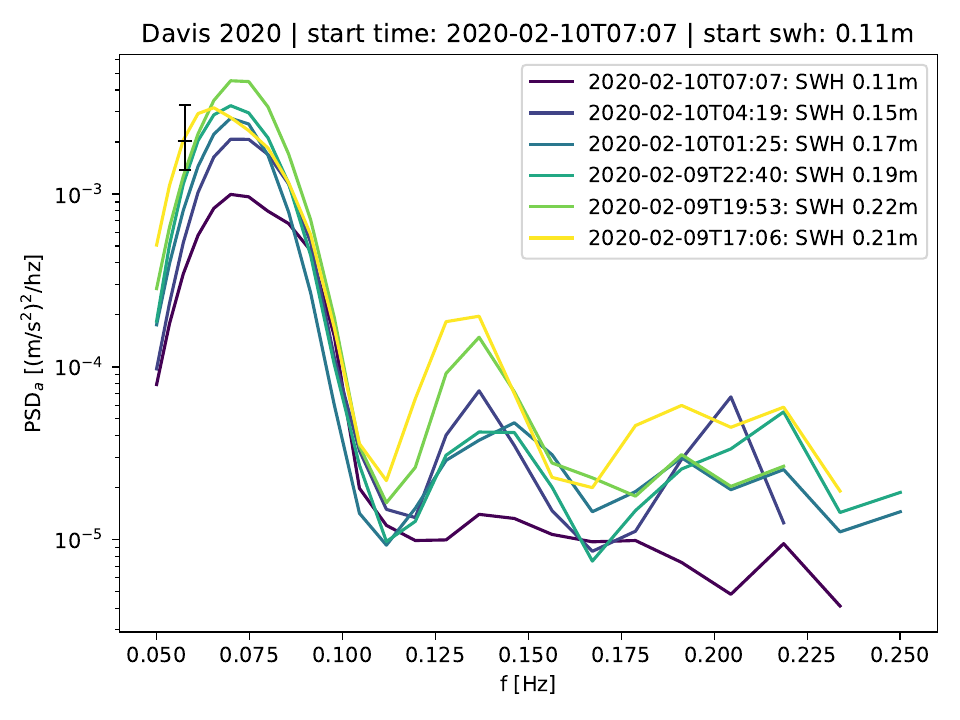}
    \includegraphics[width=0.33\linewidth]{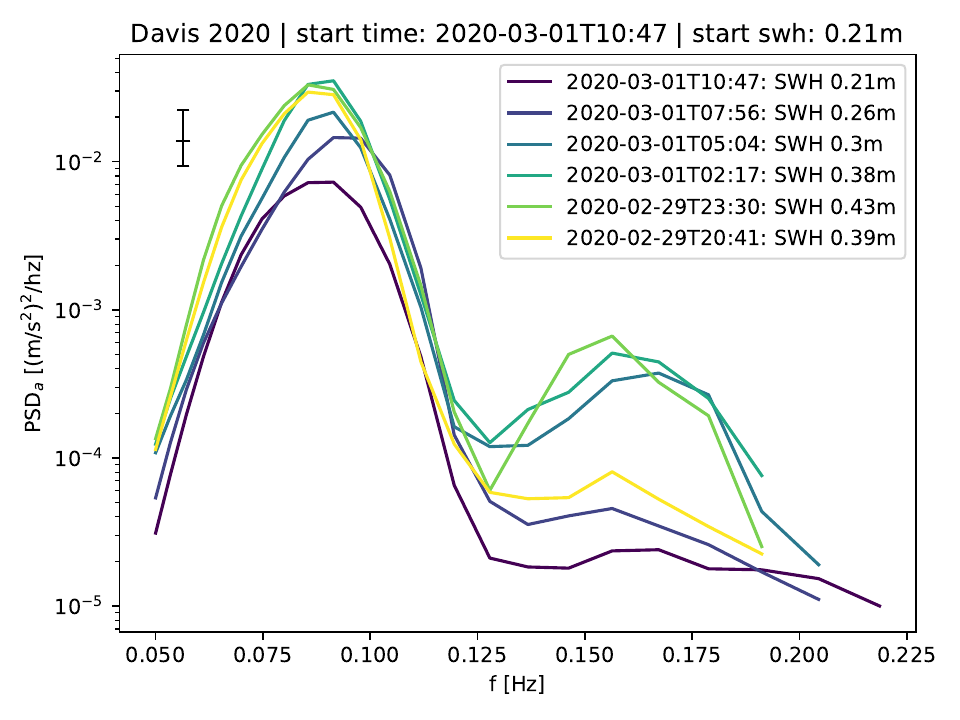}
    \includegraphics[width=0.33\linewidth]{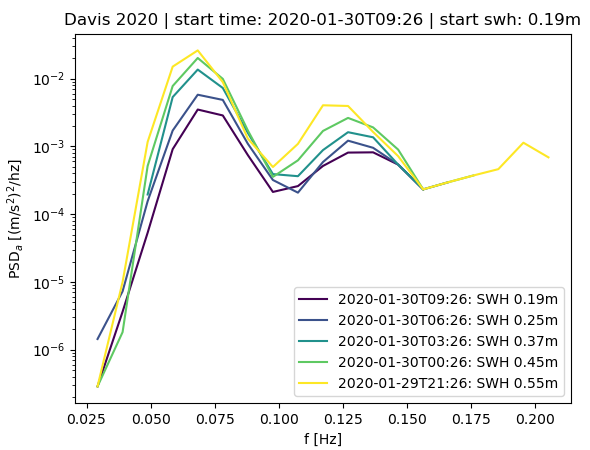}
    \includegraphics[width=0.33\linewidth]{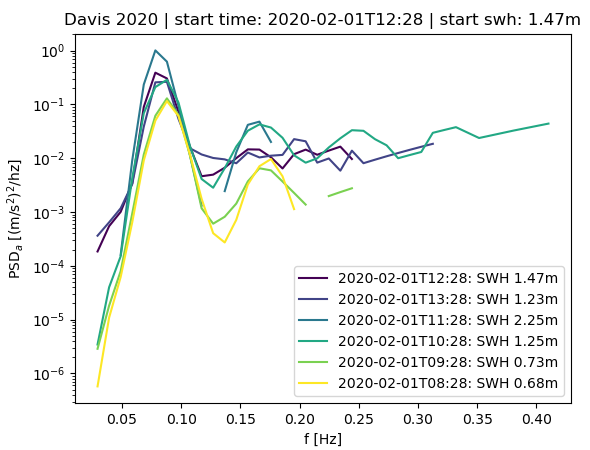}
    \includegraphics[width=0.33\linewidth]{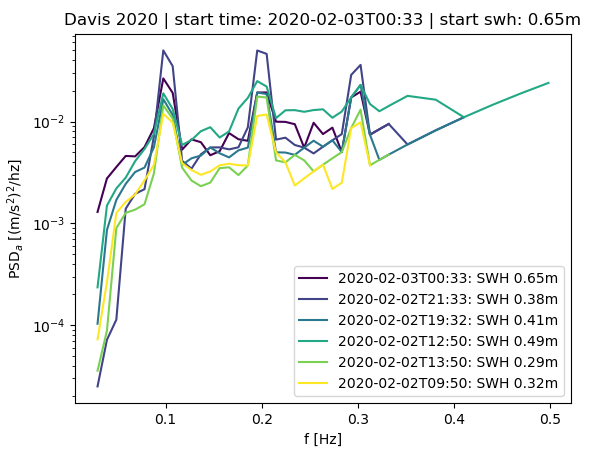}
    \caption{Examples of high-frequency peaks in the PSD of waves in ice data from Antarctica, from the Casey 2022 deployment (using OMB-v2021, first row) and the Davis 2020 deployment (using OMB-v2018, second row, and using Sofar Spotter, third row). Clear high-frequency peaks are visible, across all 3 instrument kinds, sometimes even with a third-order peak being present. Error bars (computed assuming an underlying Gaussian stochastic process) are shown as a single black cross near the top-left of the OMB-v2021 and OMB-v2018 panels; these are analytically derived from the fixed on-board Welch parameters, since only the pre-computed spectra (not the raw time series) are transmitted over Iridium. For OMB models for which the full onboard processing firmware is available open source, we compute and add these error bars as 95\% confidence intervals (on the top left of each individual plot, obtained assuming an underlying stochastic Gaussian process), which are indicated as the black whiskers. For the Sofar Spotter, which is a closed-source commercial instrument, the details of the embedded processing are not available, and we are not able to provide such error bars, though we assume they would be similar.}
    \label{fig:antarctic_psd_examples}
\end{figure}

Some wave events with clear high-frequency peaks, observed in Antarctica, are presented in Fig. \ref{fig:antarctic_psd_examples}. These data were collected by 3 different kinds of instruments: OMB-v2021 (first row), OMB-v2018 (second row), and Sofar Spotter (third row). Clear second-order high-frequency peaks are observed, and in some cases even third-order high-frequency peaks are visible.

\begin{figure}
    \centering
    \includegraphics[width=0.32\linewidth]{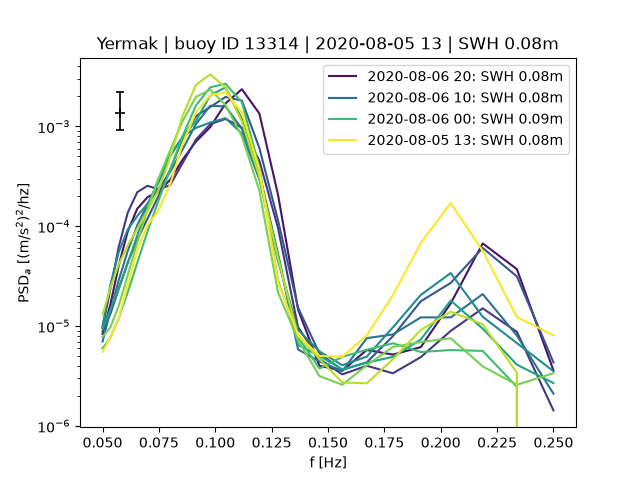}
    \includegraphics[width=0.32\linewidth]{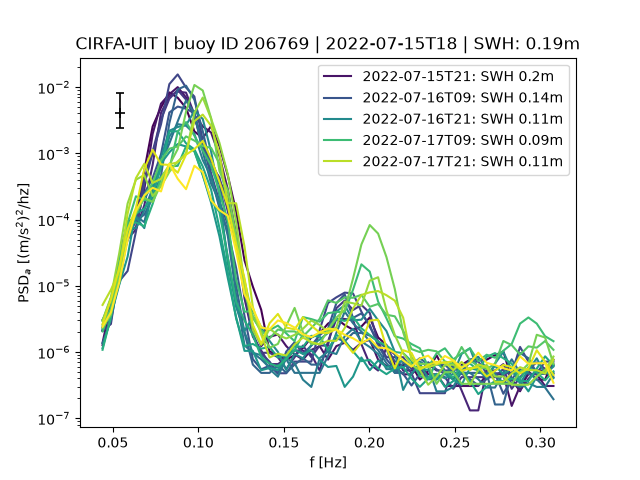}
    \includegraphics[width=0.32\linewidth]{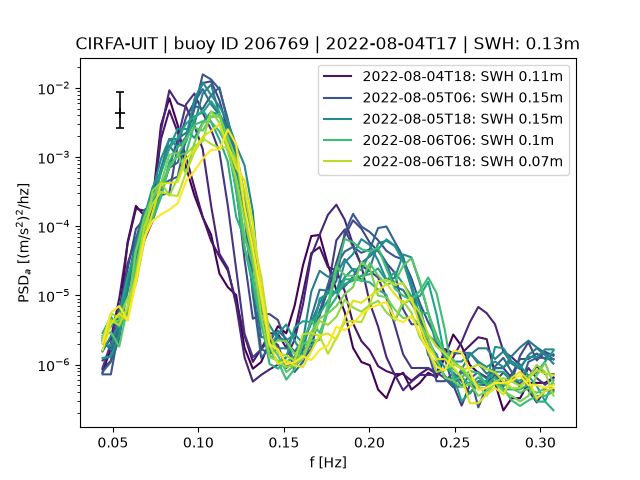}
    \includegraphics[width=0.32\linewidth]{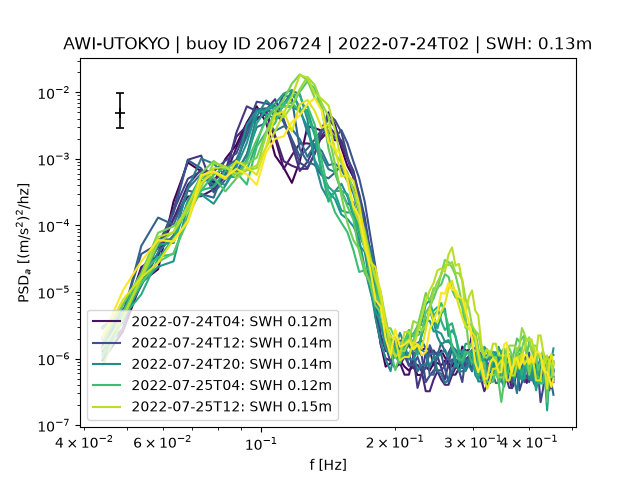}
    \includegraphics[width=0.32\linewidth]{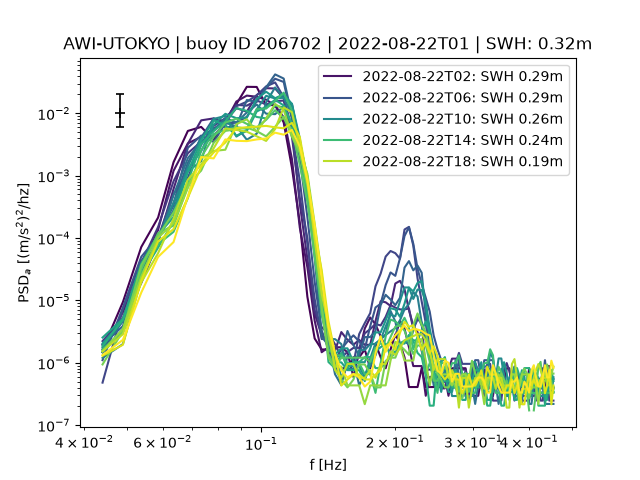}
    \includegraphics[width=0.32\linewidth]{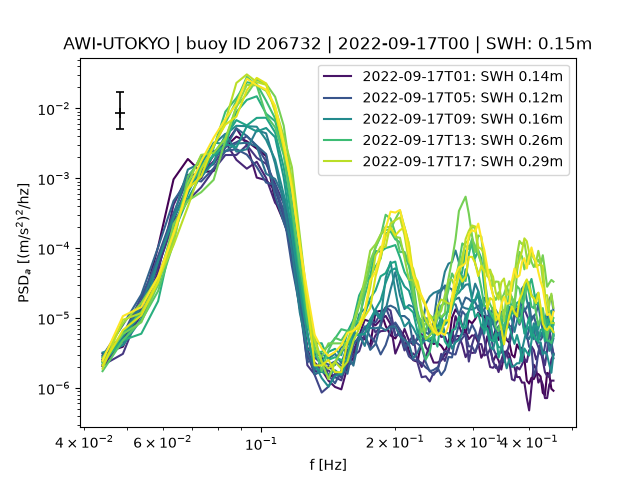}
    \caption{Examples of high-frequency peaks in the PSD of waves in ice data from the Arctic. Second, and sometimes even third order high-frequency peaks, are observed. The examples cover a range of deployment dates, locations, ice conditions, and use both OMB-v2018 and OMB-v2021. Several features that can be interpreted as possible consequences of nonlinear dynamics are visible, as described in the text. Error bars (computed assuming an underlying Gaussian stochastic process) are shown as a single black cross near the top-left of each panel; these are analytically derived from the fixed on-board Welch parameters, since only the pre-computed spectra (not the raw time series) are transmitted over Iridium. These are indicated as 95\% confidence intervals, computed under the assumption of a stochastic Gaussian process, and plotted as whiskers on the top left of each individual plot.}
    \label{fig:arctic_psd_examples}
\end{figure}

Similarly, a few similar events collected by OMB-v2021 in the Arctic are presented in Fig. \ref{fig:arctic_psd_examples}. There, too, cases with both second- and sometimes third-order high-frequency peaks are visible. Note that the AWI-UTOKYO data show the PSDs reported by three different buoys over three different time periods, so these cover several events. More events (not all reproduced here) are visible in the data. In addition to the presence of the high-frequency peaks by themselves, a number of interesting features are visible:

\begin{itemize}
    \item The strength of the high-frequency peak in the CIRFA-UIT 2022-07 plot seems to be very sensitive to the exact features of the main peak. In particular, a much higher high-frequency peak is observed for the light-green data segment, while this corresponds only to a subtle change (slightly higher peak frequency) in the main peak. This may be seen as a possible signature of strongly nonlinear mechanisms being at play.
    \item In CIRFA-UIT 2022-08, the peak frequency of the main high-frequency peak is slowly increasing over time. This is associated with a well-matched shift of the frequency of the high-frequency peak to higher frequencies. This seems to confirm the tight coupling between the main peak and the high-frequency peak, i.e., that these two are coupled and change in unison, and are not just unrelated peaks coming from unrelated phenomena that are here because of a coincidence.
    \item In AWI-UTOKYO 2022-07, it seems that the presence of the high-frequency peak can be very sensitive to the exact shape of the main peak.
    \item In AWI-UTOKYO 2022-08, the shape of the main peak remains quite similar, with a moderate reduction in amplitude occurring. However, the relative reduction in the amplitude is much greater in the high-frequency peak. Although the main peak is reduced by a factor of around O(10) between the first and last spectra (from around $4e^{-2}$ to around $5e^{-3}$), the high-frequency peak is reduced by a factor of around O(100) (from around $2e^{-4}$ to around $3e^{-6}$), in a way that is compatible with a second-order effect. A similar effect is observed in the last panel.
\end{itemize}

In addition to these observations, we also considered the PSD spectra from deployments in the MIZ in the case where small broken ice floes were encountered. These include in particular the outermost buoys in the dataset 2021-03 of \citet{rabault2023dataset}, see more descriptions in \citet{nilsen2021pc}, and the outermost buoys in the datasets of \citet{muller2025distributed,muller2025distributeddata}. There, we could not find clear high-frequency peaks similar to those reported in Fig. \ref{fig:arctic_psd_examples}. This further hints at the importance of having an ice sheet or a large enough ice floe to support significant flexural effects for observing the high-frequency peaks we report.

Since the bicoherence analysis cannot be conducted based on the 1-dimensional PSDs that are available to us alone, and we do not have enough information to confidently determine the dispersion relation and associated wave triad diagrams, we cannot assert with certainty that these high-frequency peaks are phase-locked with the main peak nor that they are compatible with the local dispersion relation to correspond to wave triads. Therefore, these observations should only be considered as circumstantial evidence for the possible existence of nonlinear mechanisms such as wave triads. However, the relatively common occurrence of these high-frequency peaks across a range of conditions, deployments, and instruments confirms that they most likely correspond to a real physical phenomenon and are not a rare one-time coincidence or technical artifact.

\section{Discussion}

\noindent \textit{Summary of results}

We have considered three kinds of in-situ observation dataset: (i) the T15 dataset, where the timeseries from several closely-located time-synchronized IMUs deployed in Arctic landfast ice are available, (ii) the Y20 dataset, where the timeseries from a single IMU-based buoy deployed in the Arctic was recovered, and (iii) a number of deployment datasets in both the Antarctic and the Arctic, where only 1-dimensional PSDs are recovered over Iridium satellite communications.

In all cases, clear high-frequency peaks can be observed in some of the data segments. Since these high-frequency peaks are clearly above the noise background, are sometimes present and sometimes not (so they are not a systematic noise in the sensors and / or processing algorithms), and these can be observed over a range of instruments (VN100-based loggers, OMB-v2018 with a VN100 IMU, OMB-v2021 with an ISM330DHCX motion sensor, Sofar Spotter using GPS Doppler data) and deployments in various sea ice conditions and locations covering both the Arctic and the Antarctic, we can be confident that these are not noise nor a one-off coincidence, but a real recurring physical signal. Moreover, these peaks are relatively narrow isolated harmonics of the main energy peak and, therefore, most likely a different mechanism from the "rollover effect" that is sometimes reported in the literature \citep{wadhams1988attenuation} and may be associated with local wind energy injection or nonlinear effects \citep{li2017rollover,Cheng2025attenuation}, though debates have arisen in the community regarding the noise levels and validity of some of the observations \citep{thomson2021spurious}. Therefore, independently of the ongoing debate around the validity and presence of the rollover effect, we believe that our present observations of harmonic peaks in waves in ice energy data indicate a distinct signature and different mechanism.

In the cases where the timeseries are obtained (T15 and Y20), a bicoherence analysis confirms that the maximum of the high-frequency peak is phase-locked with two components of the main peak. This is a necessary (but not sufficient) condition for the high-frequency peak to be the product of e.g. nonlinear wave triad interactions or some other nonlinear physical mechanism related to the main peak in the PSD.

In the one case where timeseries from several closely located and time-synchronized sensors are obtained (T15), the dispersion relation is recovered from a CSD phase-shift analysis. The results indicate that in the signal segments for which the high-frequency peak is observed in the PSD and is visible in the bicoherence analysis, clear flexural-elastic thin plate effects are present in the dispersion relation, which is compatible with the formulation of Eqn. (\ref{eqn_reduced_disprel}). This elastic-flexural component makes it possible for the dispersion relation for the waves propagating in the ice to support wave triads. We also show that the associated wave triads diagram indicates that matching triads can exist between the frequencies revealed by the bicoherence analysis.

~\\
\noindent \textit{Are these harmonics bound or free waves?}

A critical question to elucidate regarding the high-frequency spectral peaks is whether these correspond to bound or free waves (we use a similar nomenclature as, e.g., \citet{masuda1979dispersion,mitsuyasu1979dispersion,waseda2015third}). Bound waves, which canonical example is the higher-order components of the Stokes solution \citep{lamb1924hydrodynamics}, are perturbations to the first-order wave solution that are present as the higher-order consequence of nonlinear terms. These can arise either from the effect of nonlinear terms on a single wave mode (as is the case with the higher order Stokes wave components), or as the effect of nonlinear terms on a superposition of wave modes (as the presence of nonlinear terms implies that the linear superposition of two or more wave modes results in an additional nonlinear residual term). Bound waves have a relation between $k$ and $\omega$ different from the dispersion relation of usual free waves, and are caused by and travel together with the main propagating wave, materializing as "disturbances" of the main wave. Free waves, on the contrary, propagate freely and follow the dispersion relation. Based on Fig. \ref{fig:t15_disprel}, it seems most likely that the high-frequency peak observed in the T15 dataset corresponds to a free wave that follows the dispersion relation. However, we acknowledge that the signal-to-noise ratio used there is low, so offering a definite confirmation that our high-frequency spectral peaks correspond to free waves would require additional observation and confirmation. However, an additional observation can be made that strengthens the case for the waves observed being free waves: namely, the bicoherence blob in Figs. \ref{fig:t15_bicoherence} and \ref{fig:y20_bicoherence} are slightly but significantly off the diagonal - indicating, as discussed above in particular for the T15 case, that two distinct low frequencies are involved in the triad. This would also tend to rule out that the harmonic observed is a simple, single-mode Stokes bound wave, which, if this were the case, would correspond to trapped self-harmonics of each mode. Note, however, that this argument only excludes this simple self-harmonic mechanism, and does not by itself rule out the more general case of a bound wave arising from the superposition of two distinct wave modes.

The single-mode Stokes bound-wave hypothesis can, in addition, be tested quantitatively using an order-of-magnitude estimate of the wave steepness. For a weakly nonlinear Stokes wave, the amplitude of the bound second-harmonic component is $a_2 \approx (\epsilon/2) \, a_1$, where $a_1$ is the amplitude of the fundamental and $\epsilon = k_1 a_1$ is the wave steepness \citep{lamb1924hydrodynamics}. Taking the T15 segment 2 example highlighted in Figs. \ref{fig:t15_psd} and \ref{fig:t15_disprel}, the main-peak significant wave height obtained from the acceleration PSD (using $H_s = 4\sqrt{m_0}$, with $m_0$ the zeroth moment of the elevation spectrum, \citep{tucker2001waves}) is only $H_{s,\mathrm{main}} \approx 5$-$6$~cm when integrated around the peak frequency blob of about $0.18$-$0.20$~Hz. Using the best-fit flexural-elastic dispersion relation of Fig. \ref{fig:t15_disprel} ($h=0.6$~m), this corresponds to a wavelength of about $60$~m, i.e. a very mild steepness $\epsilon = k_1 a_1 \approx 0.003$. Such a steepness predicts a Stokes-bound second harmonic with a significant height of only $H_{s,2,\mathrm{pred}} \approx 0.08$-$0.10$~mm, i.e. a negligible perturbation. This is compatible with the outermost PSD (VN1005), and with segment 16, but not with the innermost PSDs of segment 2. Indeed, in contrast, directly integrating the observed high-frequency peak over the same segment, after conservatively subtracting the local noise floor estimated from the notch between the main and harmonic peaks, gives $H_{s,2,\mathrm{obs}} \approx 1.9$-$2.6$~mm for the two IMUs (VN1002 and VN1007) for which this harmonic-band signal is robustly above the local noise floor - i.e. about $22$ to $27$ times larger in amplitude, or roughly $500$ to $700$ times larger in energy, than what the Stokes bound-wave mechanism could produce for the actually observed (mild) sea state. The third IMU (VN1005, the one located closest to the ice edge) shows only a marginal high-frequency signal that is not robustly distinguishable from the local noise floor, and is therefore excluded from this quantitative estimate; this is consistent with the fact that this same IMU is also the one for which the dips associated with the triad-related energy transfer discussed above are least pronounced. Since the observed high-frequency peak is far too energetic to be explained by a single-mode Stokes bound harmonic riding on the fundamental, this quantitative argument independently corroborates, and strengthens, the free-wave interpretation discussed above. The same check can be repeated for the Y20 segment 322 example of Fig. \ref{fig:y20_psd}. In this case, no multi-sensor CSD-derived dispersion relation is available, so the open-water dispersion relation is used instead as a proxy for the wavelength at the main-peak frequency. This is not because we have evidence ruling out flexural-elastic effects in this dataset - on the contrary, for a wave triad to exist at all the dispersion relation must depart from the open-water case, so some degree of ice-induced dispersion is in fact expected if triads are well the cause for the harmonic observed - but because, at the low frequencies involved here, such a deviation is typically small. Using the T15 flexural-elastic dispersion relation of Fig. \ref{fig:t15_disprel} as a reference, the wavenumber departs from its open-water value by about only $1$\% at $0.10$~Hz, growing to $24$-$32$\% only at the higher T15 peak frequency of $0.18$-$0.20$~Hz; since the Y20 main peak sits at an even lower frequency ($\approx 0.09$~Hz), a comparably mild deviation from open water is expected. The main peak gives $H_{s,\mathrm{main}} \approx 20$~cm at a peak frequency of about $0.09$~Hz, corresponding to a wavelength of about $200$~m and a steepness $\epsilon \approx 0.003$, i.e. similarly mild as for T15. The predicted Stokes-bound second harmonic is again very small, $H_{s,2,\mathrm{pred}} \approx 0.3$~mm, while the same noise-floor-corrected integration of the observed high-frequency peak gives $H_{s,2,\mathrm{obs}} \approx 2.2$~mm, i.e. about $7$ times larger in amplitude and close to $2$ orders of magnitude larger in energy than the Stokes bound-wave prediction. This confirms, using an independent dataset and a different (open-water) proxy for the dispersion relation, that the observed harmonic is again most likely too strong to be accounted for by a simple Stokes bound wave. We also checked that this conclusion is not an artifact of the specific frequency bands chosen to integrate the main and harmonic peaks: repeating the calculation with band widths ranging from the tight core of each spectral lobe to their full visible extent changes the amplitude ratio between about $26$ and $121$ for T15 (VN1002/VN1007), and between about $6$ and $11$ for Y20 - always far above unity, so the qualitative conclusion is insensitive to this choice. In principle, this same quantitative check only requires the 1-dimensional PSD itself and does not require access to the underlying time series - the main-peak $H_s$ and peak frequency (to estimate the wavelength and steepness via the dispersion relation) and the observed harmonic-band energy can all be obtained directly from the PSD, without needing access to the underlying timeseries. As a partial, independent confirmation, we repeated this exact calculation directly from the numerical PSD values telemetered by an OMB-v2021 buoy deployed at Casey (Antarctica) in Dec. 2022 (a subset of the deployments summarized in Fig. \ref{fig:antarctic_psd_examples}). Across $24$ individual spectra spanning sea states from $H_{s,\mathrm{main}} \approx 0.07$~m to $1.1$~m, the same procedure consistently gives mild steepness values ($\epsilon \approx 0.001$-$0.03$) and observed harmonic peaks $3$ to $40$ times larger in amplitude (and about $1$ to $3$ orders of magnitude larger in energy) than the corresponding Stokes-bound prediction - the same order of magnitude as found for T15 and Y20 above. For the remaining cases, we perform a simple visual inspection of these PSDs which indicates significant wave heights of a few tens of centimeters at peak frequencies of about $0.08$-$0.13$~Hz, which, assuming an open-water dispersion relation, corresponds to steepness values of the same mild order ($\epsilon = O(10^{-3})$) as for the T15, Y20, and Casey cases discussed above; at the same time, the high-frequency peaks visible in these PSDs typically reach a comparable, and in some cases even larger, fraction of the main peak amplitude than in the T15, Y20, and Casey examples. In other words, simply observing the typical relative levels of the high-frequency peaks in these additional PSDs gives the same qualitative impression as the quantitative T15, Y20, and Casey checks above. This suggests that the same conclusion holds: the observed high-frequency peaks are too energetic to be explained by a simple single-mode Stokes bound wave, or by a mechanism of similar magnitude. This is also consistent with the fact that these harmonics are sometimes observed, and sometimes not - if they would come from a Stokes-like bound wave, they should by contrast be consistently present or absent given similar main PSD peak characteristics.

In addition to these arguments, the distinction between the properties of bound versus free waves points to additional considerations. Namely, a bound wave naturally exhibits phase lock with the carrier wave (and associated high bicoherence), and in the case of a Stokes bound wave for example the biphase between the (self-interacting) fundamental and the harmonic is exactly $0$. However, a free wave harmonic has generally no a priori reason to be phase locked at a specific relative phase with the associated fundamental waves at a given location and time. This, together with detuning mechanisms, results in different direction for the energy transfer between the fundamental modes and free wave harmonic in Eqn. (2.5) of \citet{marchenko1999stability} and similarly Eqn. (3.4) of \citet{pierce2024sum} over space and time, which is also visible in the spatial variability in particular the existence of nodes and antinodes as highlighted in e.g. Fig. 5 of \citet{pierce2024sum}. Generally, a high value of the bicoherence between a free wave harmonic and the fundamental modes is the indication that actual physical coupling is taking place locally between them, and that energy is being exchanged between the fundamental and harmonic modes over time at the location considered, with the direction and amount of energy transfered depending, among others, on the relative phase of the coupling between the modes at this location. These considerations on phase and energy transfer can, therefore, be used to further investigate (and distinguish between) the nature of the harmonics we report.

In particular, it is easy to observe from the Stokes wave solution that the biphase between the (self-interacting) fundamental mode and its Stokes harmonic is $0$ as mentioned above. However, one does not expect such a $0$-biphase phase lock to hold generally in the case of free harmonics and free wave triads, and a non-zero biphase is even indicative of what energy transfers are taking place. Here, we choose to illustrate this by starting from the Manley-Rowe (MR) like equations of \citet{pierce2024sum}, which describe the energy transfer between three resonant modes as the waves propagate in space, assuming that conditions are stationary in time. \citet{marchenko1999stability} derives very similar MR-like equations, though focusing on the evolution in space and time of a single set of waves. \citet{pierce2024sum} only considers the case where the three resonant waves are propagating in the same direction; while this most likely is not strictly the case here as visible from our triad diagrams discussed above, this allows for them to derive more specific expressions than the general complex sums of \citet{marchenko1999stability}. However, MR-like equations have a general form arising from the nature of 3-wave interaction with quadratic coupling; this means that one should remember that while the exact coefficients and phases appearing for three non-colinear waves will be, strictly speaking, modified from what we derive here as their relative angles increase, the general form of the equations will be similar. Therefore, taking into account the angle between waves involved in a triad, or similarly other effects ignored by \citet{pierce2024sum} such as viscosity and attenuation, will modify the exact numerical values obtained below but not change the fundamental results.

Keeping these limitations in mind, we start from Eqn. (3.4c) of \citet{pierce2024sum}, using a time-advection derivative rather than a time partial derivative (as is done in the equivalent but slightly more complex form written in Eqn. (2.5) of \citet{marchenko1999stability}), and remembering that all 3 waves are assumed to be colinear (which simplifies the equations and leads to closed forms, as shown in \citet{pierce2024sum}):

\begin{equation}
  \frac{d a_3}{d T} = - i \frac{\omega_1 \omega_2}{2 |C_{p3}|} a_1 a_2,
\end{equation}

\noindent where $a_i(t)$ is the complex amplitude of mode $i$ at the frequency $f_i$, $\frac{d}{d T} = \frac{\partial}{\partial \bar{t}} + c_{g3} \frac{\partial}{\partial \bar{x}}$ is the material derivative following the wave at its group speed, $c_{g3} = \frac{\partial \omega}{\partial k}(f=f_3)$ the group speed of the mode considered, $c_{p3} = \frac{\omega}{k}(f=f_3)$ the phase speed, $\bar{t}$ and $\bar{x}$ are slow-varying normalized time and position variables as defined in \citet{pierce2024sum}, and $C_{p3}$ a numerical coefficient (which in the 3D case of \citet{marchenko1999stability} arises from the complex summation written therein). In the formalism of \citet{pierce2024sum}, the $a_i$ are equivalent to the complex Fourier coefficients used in frequency decomposition over a set of modes.

From there, one can easily derive (see Appendix C for the actual derivation) that the amplitude of the harmonic follows:

\begin{equation}
\frac{d |a_3|}{dT} = \frac{\omega_1 \omega_2}{2 |C_{p3}|} |a_1| |a_2| \sin(\phi_1 + \phi_2 - \phi_3),
\end{equation}

\noindent where $\phi_i$ is the phase associated with the mode $a_i$. This means that the biphase (phase of the bispectrum) \mbox{$\Delta~\phi~=~\phi_1+\phi_2-\phi_3$} indicates the direction, and contributes to the magnitude, of the energy transfer from modes 1 and 2 to mode 3: the energy transfer to the mode 3 is maximum when $\Delta \phi = \pi / 2$, zero when it is equal to $0$ or $\pi$, and maximum away (extracting from) mode 3 when equal to $-\pi / 2$. Remember that these results are only valid in the idealized case of inviscid, non-attenuated, zero viscosity, colinear 3-wave triads, so that the exact details and values may be slightly (for small deviations from these approximations) different when all these effects play a role, which is the case in real world situations. Due to several detuning phenomena that happen on top of the triad resonance conditions (such as the nonlinear amplitude-dependent dispersivity of the medium, the effect of viscous damping, and other real world effects), $\Delta \phi$ can in real world situations vary with space, creating node and antinode patterns for the slowly varying (in space) wave modes energy distribution. In our present case, we can for example compute the biphase at the location of IMU VN1002 from the T15 dataset (similar results are obtained with VN1007), which is the case for which we have the best data. The resulting biphase at the peak of the bicoherence is equal to $135$ degrees, with error bars putting the actual value somewhere between $125$ and $155$ degrees following a 95\% confidence interval. The biphase over the whole high bicoherence blob is generally similar in value, as shown in Appendix D. This is a further indication that the bicoherence results observed are not consistent with a Stokes like harmonic, while they are consistent with a forced wave triad harmonic (typically if the approximations relied on above are acceptable, indicating energy injection into the harmonic from the two lower frequencies), in good agreement with the evolution of the PSD from VN1005 to VN1002 and VN1007. In the other case where we can compute the bispectrum, Y20, the biphase is typically around $140$ degrees though significant variability is observed, leading to similar conclusions (we cannot determine by other means if the IMU is located at a location where energy transfer is into or out of the harmonic as we do with T15, but the fact that the phase is significantly different from $0$ tends to rule out a Stokes-like bound harmonic).

Another possible mechanism worth considering is whether the observed bicoherence signature could arise from an interaction between a single incident wave and a heterogeneous propagation medium (e.g. a spatially varying ice thickness and/or elastic modulus, $h=h(x)$ and/or $E=E(x)$), rather than from a genuine nonlinear three-wave interaction. This appears unlikely for both the T15 and Y20 datasets. The defining feature of the bicoherence used here is that the high-frequency component is phase-locked with a specific relative phase difference between two distinct lower-frequency components simultaneously - i.e., it captures a relation involving three, not one, frequency components. This is precisely the signature expected from a nonlinear three-wave interaction, and would not be expected from the (linear) scattering of a single incident wave off spatial heterogeneities in the propagation medium, however strong these might be. A heterogeneous propagation medium would, nonetheless, be expected to modulate the details of the energy transfer between the three waves involved in a resonant triad, once the triad resonance condition is otherwise satisfied; this can be seen, for example, in the equations discussed in \citet{marchenko1999stability,marchenko1999stability2}, where spatial gradients in ice properties introduce additional terms involving derivatives of the group velocity coefficients, without altering the fundamentally nonlinear, three-wave nature of the interaction. Quantifying this modulation in detail would require a dedicated nonlinear (e.g. potential-flow) numerical treatment, together with a precise knowledge of the spatial distribution of ice thickness and elastic properties (for example from ice radar or electromagnetic sounding measurements), which goes beyond the scope of the present manuscript, whose focus is on demonstrating the existence of such nonlinear interactions from in-situ data rather than on quantifying the associated energy transfers in detail. It is also worth noting that an analogy with wave shoaling, whereby a decrease in group speed in shallower water is compensated by an increase in wave height so as to conserve the energy flux, would not support this alternative mechanism either: as waves propagate from open water into thicker sea ice, both the phase and the group speed increase, if anything, rather than decrease (see e.g. Fig. 17 of \citet{rabault2016measurements}), so that any such linear effect would tend to reduce, rather than amplify, the high-frequency wave component, and even more so at increasing wave frequency.

These results taken together can be seen as a strong hint that nonlinear dynamics are a likely cause of the high-frequency peaks observed and that triads are a strong candidate for the underlying energy-transfer mechanism. However, we want to highlight that our results indicate that the necessary conditions for triads are fulfilled, but that these are not strictly speaking sufficient conditions / observations to assert with complete certainty that triads are indeed the underlying mechanism causing the energy transfers. To bring certainty to this matter, additional measurements designed specifically to test this hypothesis would be needed, as suggested below.

An additional observation is that wave directionality seems to play a role. While the frequency and wavenumber triad conditions provide necessary conditions for triads to be possible, these say nothing about the strength and direction of the energy transfers between the three frequencies involved. In particular, the extraction of energy from the two low-frequency components and transfer to the high-frequency one may only happen for some particular frequencies and relative orientation conditions. For example, in the case of the T15 dataset, we note that the candidate triad occurs for wave components 1 and 2 that have a significant relative angle, keeping in mind that this is observed deep in a narrow fjord system (and even changing a bit the parameters involved, significant angle is observed, as long as one does not go to the extremes of the parameters that can sustain triads). Similarly, a quick investigation of when high-frequency peaks are obtained in Antarctica seems to indicate that these happen mostly relatively close to open water (this is based on a visualization of the sea ice concentration maps in the area, not reproduced here). This may be consistent with the fact that, as waves propagate through the ice, the directional spectrum becomes more narrow (since components going sideways have longer effective propagation distances and are, therefore, more strongly attenuated). However, we consider that an exhaustive and detailed meta-studies of when and where these high-frequency peaks occur in all OMB and other wave in ice datasets openly available, and a comparison to incoming wave spectra and sea ice conditions, would be a major effort that is out-of-scope for the present manuscript.

While our observations seem to be compatible with the existence of nonlinear wave interaction and energy transfers between frequency modes through a wave triad mechanism, other mechanisms could also play a role and / or explain our observations. For example, \citet{zhao2017gap} suggests that arrays of adjacent floating bodies (in our case, ice floes) can lead to nonlinear resonance and create high-frequency peaks in PSDs. However, this seems unlikely to be the cause of the observations in T15. Indeed, in this case, the high-frequency peak is observed when a flexural dispersion relation is present, i.e. when the ice is unbroken, while it disappears when the flexural component disappears, i.e. the floes get broken. This is the opposite of what could be expected from the floes array resonance mechanism discussed in \citet{zhao2017gap}. Other possible mechanisms could involve, for example, edge waves, which can also participate in sustaining nonlinear resonant energy transfers \citep{marchenko1999parametric}. Note that, to make the matter even more complex, nonlinear energy transfers through resonant triads could even happen across different kinds of waves \citep{phillips1981wave} (edge waves being a possible candidate). There may also be other mechanisms that we have not yet imagined and that could also result in these high-frequency peaks.

~\\
\noindent \textit{Possible future experiments}

In order to answer these questions with certainty, a purposely-designed set of experiments can be envisioned.

\begin{itemize}
    \item Ideally, in the field, one should perform detailed time-resolved and time-synchronized wave acceleration measurements in a wide area of sea ice, in order to be able to resolve the full dynamics of the two-dimensional ice sheet. This is a realistic attempt anno 2025, thanks to the low cost and accuracy of modern motion sensors such as the ISM330DHCX used in the OMB-v2021 (cost O(20 USD)), and the time accuracy provided by GPS Pulse Per Second (PPS) information (UTC second start can be determined with a jitter of at most O(10 ns) using GPS chips that cost O(30 USD)). This was used, for example, in \citet{voermans2023estimating}, which designed and deployed a low-cost open source logger that could reach microsecond accuracy in signal acquisition timestamps and logging rates of several kHz on a SD card, using a low-cost micro controller. With proper engineering, an open source SD-based wave in ice logger can be built with high-accuracy wave measurements and (sub-) microsecond timing accuracy for O(150 USD) including all material costs, as illustrated by \citet{rabault2026openwavelogger}. This means that the hardware for a large-scale measurement campaign with 200 such loggers deployed as an array on the sea ice would cost O(30k USD), well within reach of a small- to medium-sized scientific project. This would enable a detailed analysis of the in-situ dispersion relation, spatial and spectral distributions of the energy, and of the effect of directionality.
    \item In addition, detailed analytical calculations, as well as numerical simulations of the waves in ice equations using a 3-dimensional nonlinear solver and considering the effect of the full directional spectrum, could help to understand the energy transfers at stake \citep{constantin2018nonlinear}.
    \item Finally, it could be possible to investigate these phenomena in a laboratory experiment. However, this may be challenging due to (i) the need to properly scale the physical phenomena happening, which can result in several non-dimensional quantities conflicting with each other as it has previously been observed in the study of waves in ice attenuation \citep{rabault2019experiments}, (ii) the likely need to consider fully 2-dimensional directional wave spectra, (iii) inherent challenges related to reflections and standing modes in a wave tank, especially when the effects to measure are subtle higher order effects that can easily be hidden by larger systematic errors induced by wave reflections, imperfect paddle-matching conditions, and finite-length effects of either the wavetank or the wave packets generated, as highlighted previously e.g. in \citet{rabault2016ptv,marchenko2021laboratory}.
\end{itemize}

Before some additional, purposely gathered data are available, a natural extension to the present work is to perform an exhaustive meta-study on all available OMB and similar waves in ice data. In particular, it would be interesting to systematically compare when high-frequency peaks are observed with both the position of the buoys in the MIZ, the incoming wave conditions, the atmospheric conditions, and satellite imaging of the local sea ice configuration, to attempt to reveal critical factors that cause these high-frequency peaks to be present or absent. We consider that such a study would require a major effort to be as representative as possible and to consider as many datasets as possible, and that it is beyond the scope of the present manuscript. 

~\\
\noindent \textit{Can the harmonics be a significant contribution to the energy attenuation budget?}

An interesting question is the impact and applicability of the present findings. While the present analysis gives us confidence that the spectral high-frequency peaks are real and may likely be related to interesting nonlinear dynamics, are these a physical and mathematical curiosity without much practical value, or are they important for sea ice dynamics or can be otherwise useful? Naturally, a first possible intrinsic impact is, as discussed above (and quantified below), the possible contribution of triads to indirectly increasing the attenuation rate of waves in ice. Another possible effect on the dynamics of the sea ice is that these may contribute to ice breakup, as discussed in \citet{pierce2024sum}. From a different viewpoint, considering a remote sensing perspective, if these high-frequency peaks could be reliably observed from satellite data, then they could possibly give information about the existence of flexural stresses in the ice and, hence, about the effective properties of the sea ice (in particular, through the information carried about the value of the $B$ factor in Eq. (\ref{eqn_reduced_disprel})). However, it remains to be seen whether these high-frequency peaks can be visible in satellite images and how much information can be inferred back.

We may attempt to determine how much additional energy dissipation can be induced by the nonlinear energy transfer to the high-frequency peak and the dissipation there. To do so, we can perform some order-of-magnitude calculations, assuming the following:

\begin{itemize}
    \item We work with a time-stationarity approximation: i.e., we assume that, averaged over a time O(20 minutes) corresponding to the duration over which the spectra presented above are computed, the spectra can be taken as a representative state of the wave energy distribution.
    \item We neglect variability in space, and consider that the high-frequency spectral peaks observed are in a state of local energy balance (effectively neglecting gradients in the wave energy fluxes). Effectively, this means that the estimates we consider are valid at the location of the sensors being deployed, but that owing to the node-and-antinode patterns that are characteristic of triad-generated harmonics \citep{pierce2024sum}, there may be significant variability in space. Though this variability will average over an area large enough, it is impossible to know if the sensors are sampling a representative or an extreme range of conditions - the analysis below assumes that the sensors measure conditions representative of the average conditions - a sort of "Copernican principle" that a random sample is likely to not be any special.
    \item We consider that two mechanisms are involved in sustaining the high-frequency peak and determining its level: i) a net nonlinear energy transfer $S_{\mathrm{nl}}$ from the main peak to the high-frequency peak, and ii) wave energy dissipation $S_{\mathrm{diss},\mathrm{high-freq}}$ associated with the presence of the high-frequency peak.
\end{itemize}

Under these assumptions, the nonlinear energy transfer to the high-frequency peak and the dissipation there are in balance: $S_{\mathrm{nl}} = S_{\mathrm{diss},\mathrm{high-freq}}$. Therefore, estimating $S_{\mathrm{diss},\mathrm{high-freq}}$ provides an estimate for $S_{\mathrm{nl}}$. As a consequence, estimating the relative importance between $S_{\mathrm{diss},\mathrm{main-peak}}$ (the energy dissipation at the main peak) and $S_{\mathrm{diss},\mathrm{high-freq}}$ provides an indication of the fraction of the wave in ice attenuation in the main peak that comes from the nonlinear energy transfer and dissipation at higher frequency. To estimate this relative amount of energy dissipation between the main and the higher-frequency peaks, we use the following hypothesis:

\begin{itemize}
    \item We assume that the wave energy spatial dissipation rate is similar to what is derived by \citet{yu2022}, in particular their Fig. 4.b which is based on a large set of test data: $k_i \propto f^{4.5}$.
    \item The temporal dissipation rate at a fixed location can be calculated as: $\beta = c_g k_i$ \citep{gaster1962note,sutherland2017attenuation}; with deep water $c_g \propto f^{-1}$, this leads to $\beta \propto f^{3.5}$ (we use the open-water group velocity here for simplicity: the flexural-gravity deviation, while important to enable triad resonance, is small to moderate in relative terms for the wave propagation characteristics themselves). Therefore, at a fixed location, the temporal dissipation rate ratio between the energy dissipation in the high-frequency peak and the main peak is $\mathrm{DR}_{2f} / \mathrm{DR}_{f} = 2^{3.5}  \mathrm{PSD}_{\eta, 2f} / \mathrm{PSD}_{\eta, f}$, where $\mathrm{DR}_f$ is the temporal dissipation rate at frequency $f$.
    \item $\mathrm{PSD}_{\eta} \propto f^{-4} \mathrm{PSD}_{\mathrm{acc}}$, and we are looking at $\mathrm{PSD}_{\mathrm{acc}}$ in our graphs, while the scaling formula for $k_i$ applies to $\mathrm{PSD}_{\eta}$; therefore, $\mathrm{DR}_{2f} / \mathrm{DR}_{f} = (2^{3.5} / 2^4) \mathrm{PSD}_{\mathrm{acc}, 2f} / \mathrm{PSD}_{\mathrm{acc} f} \approx 0.7 \mathrm{PSD}_{\mathrm{acc}, 2f} / \mathrm{PSD}_{\mathrm{acc}, f} = 0.7 R_{f, 2f}$, where \newline
    $R_{f, 2f} = \mathrm{PSD}_{acc, 2f} / \mathrm{PSD}_{\mathrm{acc}, f}$ .
\end{itemize}

Taking into account the PSD plots presented in the figures above (Figs. \ref{fig:t15_psd},\ref{fig:y20_psd},\ref{fig:antarctic_psd_examples},\ref{fig:arctic_psd_examples}), $R_{f, 2f}$ shows quite a large degree of variability. For some of the PSDs in the Antarctic $R_{f, 2f} \approx O(1)$ (case "A1"); for T15 $R_{f, 2f} \approx O(0.1)$ (case T15); for most other cases, we generally observe $R_{f, 2f} \in{[O(0.1),O(0.01)]}$ (case "O"). Therefore, our order-of-magnitude estimates based on the stationarity approximation indicate that damping in the high-frequency peak may represent up to 70\%  of the intrinsic energy dissipation for the main peak in the case "A1", 7\% in the case T15, and 7\% to 0.7\% in the "O" cases. This indicates that this mechanism may, in certain circumstances, play a significant role (nearly doubling the amount of energy dissipation in the case "A1"), but that it may also in many cases be only a minor contribution to the total wave energy dissipation. However, note that these are only very coarse order-of-magnitude estimates. For example, we do not take into account the spectral width of the main vs. high-frequency peaks. Moreover, the high-frequency peak dissipation effect may already be somehow present in the wave in ice empirical dissipation laws that are fitted to observations, in which case the exponent in the frequency dependence of the energy dissipation may be biased. Therefore, additional and more detailed investigation is needed to provide a definitive conclusion, and one may need to consider a systematic, representative, large-scale, and well-sampled dataset to cover all typical cases of ice and wave conditions.

\section{Conclusion}

We show that high-frequency spectral peaks can be reliably observed in the 1-dimensional Power Spectral Density (PSD) of waves in ice buoys. Although these are only intermittently observed, we show that they occasionally occur in a wide range of deployments sampling various sea ice conditions using several types of instruments. In the cases where in-situ timeseries and not just 1-dimensional PSDs are obtained, we show that these high-frequency peaks are associated with a high value of the (self-) bicoherence. In a specific case where several sensors are placed adjacent to each other on the sea ice, we show that the empirical dispersion relation obtained from a cross-spectral density (CSD) analysis contains significant flexural effects, which makes it compatible with the existence of wave triads that match the frequencies of the high-frequency peak and the bicoherence peak.

Our results indicate that a strong set of concomitant evidence, which are compatible with each other and mutually reinforcing, suggests that waves in ice high-frequency spectral energy peaks are a relatively common mechanism and may be associated with wave triad interactions. This, in turn, can extract energy from the main spectral energy peak, and re-inject this energy into a higher-frequency, more heavily damped peak. As a consequence, this mechanism can introduce a significant contribution to the wave in ice dissipation.

Although it seems very unlikely for our observations to be the result of a coincidence, the results that can be derived from our data are all necessary but strictly speaking not sufficient conditions to establish that wave triad interaction is truly the underlying mechanism causing these high-frequency peaks. Therefore, further investigation is necessary to definitively confirm that these high-frequency peaks are induced by wave triads that cause nonlinear energy transfer from the main PSD peak to the high-frequency peak(s), and quantify the energy transfers in detail. We suggest that such further investigations can be based on a combination of more detailed and advanced field measurements, numerical studies, and laboratory experiments.

\section{Use of Artificial Intelligence tools in the writing of this manuscript}

Artificial intelligence tools, in particular Large Language Models (LLMs), were used for the writing of this manuscript. In the first version of the manuscript, LLMs were exclusively used to help write the manuscript text. In the second version of the manuscript, LLMs and Agentic AI were also used to help extend the code used to generate the figures. The scientific content of the work was the product of the work of the authors, and the authors reviewed and quality controlled each contribution from the LLMs.

\section*{Appendix A: open data and open code}

The data used in this manuscript are openly available.

\begin{itemize}
    \item The VN100 data from the Tempelfjorden 2015 expedition are released openly together with this article at the following URL: [will be made available upon peer-reviewed publication of the manuscript].
    \item The VN100 data from the Yermak 2020 expedition are openly available at the following URL, as described in \citet{dreyer2024direct}: \url{https://github.com/larswd/MIZ_Floe_collisions_archive} .
    \item The OMB data transmitted over Iridium, are already openly available at the following URLs \citep{rabault2023dataset,rabault2024position}: \url{https://github.com/jerabaul29/data_release_sea_ice_drift_waves_in_ice_marginal_ice_zone_2022} and \url{https://github.com/jerabaul29/2024_OpenMetBuoy_data_release_MarginalIceZone_SeaIce_OpenOcean}.
\end{itemize}

Some of the code used in this manuscript is openly available at the following URL: [will be made available upon peer-reviewed publication of the manuscript]. We invite readers who have comments or are willing to interact with us to do so through the Github issue tracker of the corresponding repository.

\begin{figure}
    \centering
    \includegraphics[width=0.95\linewidth]{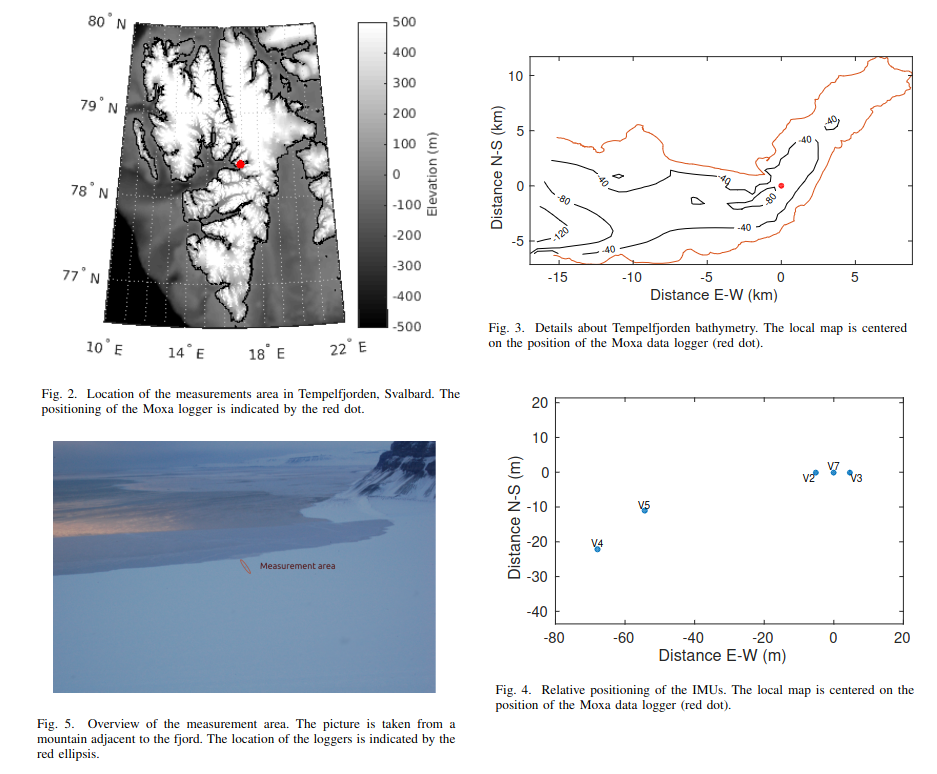}
    \caption{Illustration of the location and conditions in which the deployment T15 took place. The different panels are reproduction and collage of the figures shown in the postprint openly available at \url{https://www.researchgate.net/publication/305339430_Measurements_of_Waves_in_Landfast_Ice_Using_Inertial_Motion_Units} (accessed 2026-09-28), and the figure numbers visible with each caption correspond to this manuscript.}
    \label{fig_T15_conditions}
\end{figure}

We summarize more details about the T15 dataset, which is the most important and complete here and for which the most information about in-situ conditions is available. For the other deployments, the reader is referred to the links provided above, and significantly less information is available, since the buoys drifted over time and sea ice conditions may have changed significantly since deployment took place. The Tempelfjorden 2015 deployment (T15) took place on landfast ice in Tempelfjorden, Svalbard. The sea ice thickness was about 60-80 cm depending on the location. The deployment took place some distance away the ice edge, due to safety rules that forbade going all the way to the ice edge. The best summary of the conditions is provided in \citep{rabault2017measurements}. We reproduce the key figures from the openly available postprint available at \url{https://www.researchgate.net/publication/305339430_Measurements_of_Waves_in_Landfast_Ice_Using_Inertial_Motion_Units} in Fig. \ref{fig_T15_conditions}.

\section*{Appendix B: triad diagram conventions used in this manuscript}

Triad diagrams are used to visualize the sets of wave vectors $(\bold{k}_1, \bold{k}_2, \bold{k}_3)$ that satisfy the triad relations Eqn. (\ref{eqn_triads}). There are different conventions that can be used to plot a triad diagram. Here, we plot the longest wave vector ($\bold{k}_3$ given our conventions and the quadrant of the plane considered) along the X-axis of the plot, starting at the origin. The locus of the tip of the vectors $\bold{k}_1$ that follow these equations is plotted as a curve, which is labeled by the value of $\bold{k}_3$. This locus is a curve that depends on both the choice of $\bold{k}_3$, and the exact details of the dispersion relation. By construction of the diagram, the relation $\bold{k}_1 + \bold{k}_2 = \bold{k}_3$ is enforced. The other condition ($\omega_1 + \omega_2 = \omega_3$) is enforced by numerically determining the suitable values of $\bold{k}_1$ given the dispersion relation considered. Hence, varying $\bold{k}_3$ or the effective flexural stiffness used in the dispersion relation generates families of curves representing the corresponding loci for $\bold{k}_1$, as visible in the figures above.

\begin{figure}
    \centering
    \includegraphics[width=0.60\linewidth]{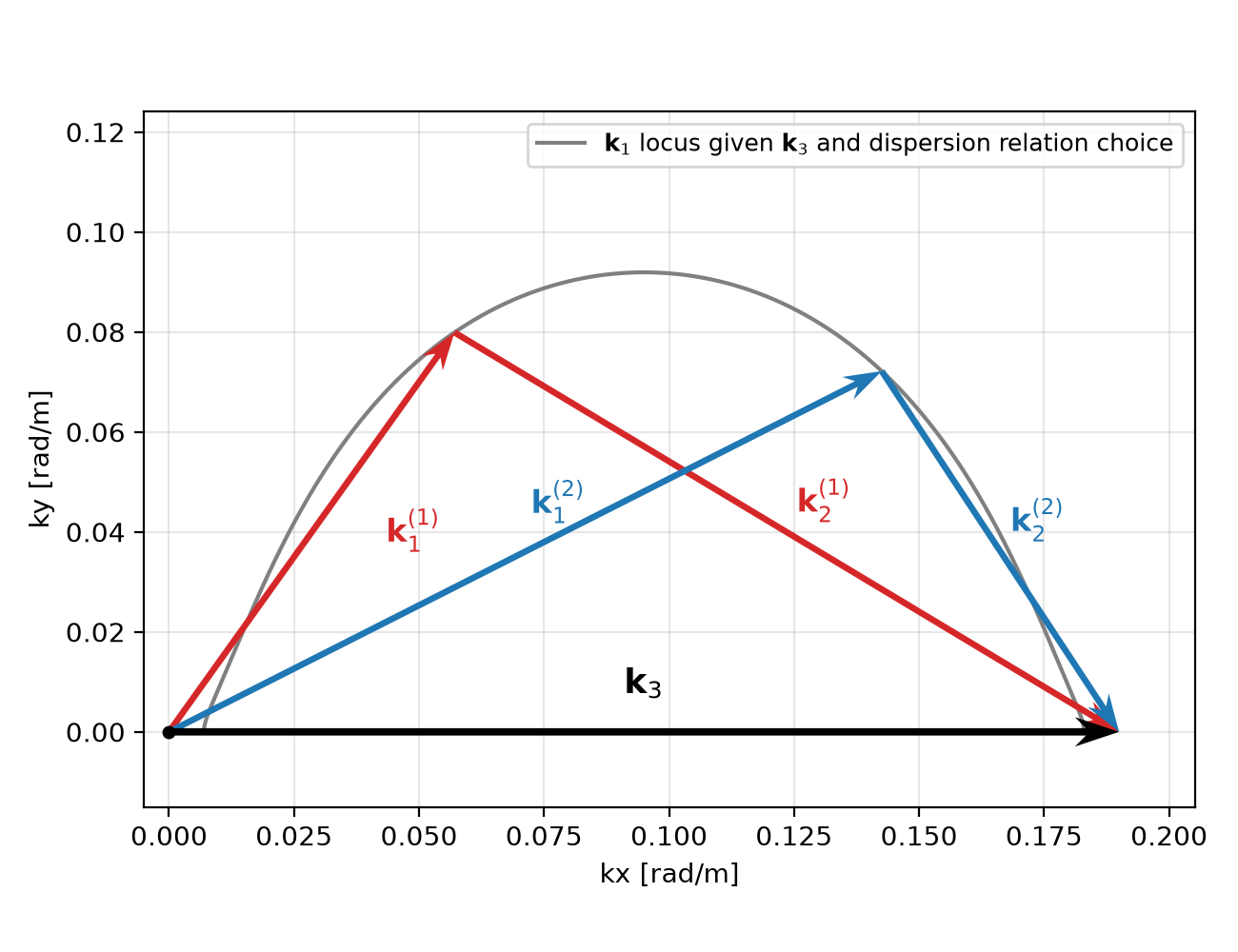}
    \caption{Schematic illustration of the conventions used in the triad diagrams presented in this manuscript.}
    \label{fig_illustrate_triads}
\end{figure}

This is illustrated by Fig. \ref{fig_illustrate_triads}, where we plot two discrete examples ($(\bold{k}_1^{(1)}, \bold{k}_2^{(1)}, \bold{k}_3)$ and $(\bold{k}_1^{(2)}, \bold{k}_2^{(2)}, \bold{k}_3)$) of triad conditions, out of the whole family of triads where the third wave vector is equal to the specific $\bold{k}_3$ chose, which is described by the grey locus curve.

%

\section*{Appendix C: derivation for the harmonic mode amplitude growth equation}

We start from the equation used in the discussion:

\begin{equation}
  \frac{d a_3}{d T} = - i \frac{\omega_1 \omega_2}{2 |C_{p3}|} a_1 a_2,
\end{equation}

We express each mode as a real amplitude and phase, yielding:

\begin{equation}
\frac{d}{dT}(|a_3|e^{i \phi_3}) = -i \frac{\omega_1 \omega_2}{2 |C_{p3}|} |a_1| |a_2| e^{i(\phi_1 + \phi_2)}.
\end{equation}

Expanding the left term:

\begin{equation}
\frac{d}{dT}(|a_3|e^{i \phi_3}) = [\frac{d |a_3|}{dT} + i |a_3| \frac{d \phi_3}{dT}] e^{i \phi_3},
\end{equation}

\noindent one gets:

\begin{equation}
\frac{d |a_3|}{dT} + i |a_3| \frac{d \phi_3}{dT} = -i \frac{\omega_1 \omega_2}{2 |C_{p3}|} |a_1| |a_2| e^{i(\phi_1 + \phi_2 - \phi_3)}.
\end{equation}

Taking the real part on both sides yields the result used in the discussion:

\begin{equation}
\frac{d |a_3|}{dT} = \frac{\omega_1 \omega_2}{2 |C_{p3}|} |a_1| |a_2| \operatorname{Re}\!\left[-i\, e^{i(\phi_1 + \phi_2 - \phi_3)}\right] = \frac{\omega_1 \omega_2}{2 |C_{p3}|} |a_1| |a_2| \sin(\phi_1 + \phi_2 - \phi_3).
\end{equation}

\section*{Appendix D: plots of the biphase}

\begin{figure}
    \centering
    \includegraphics[width=0.78\linewidth]{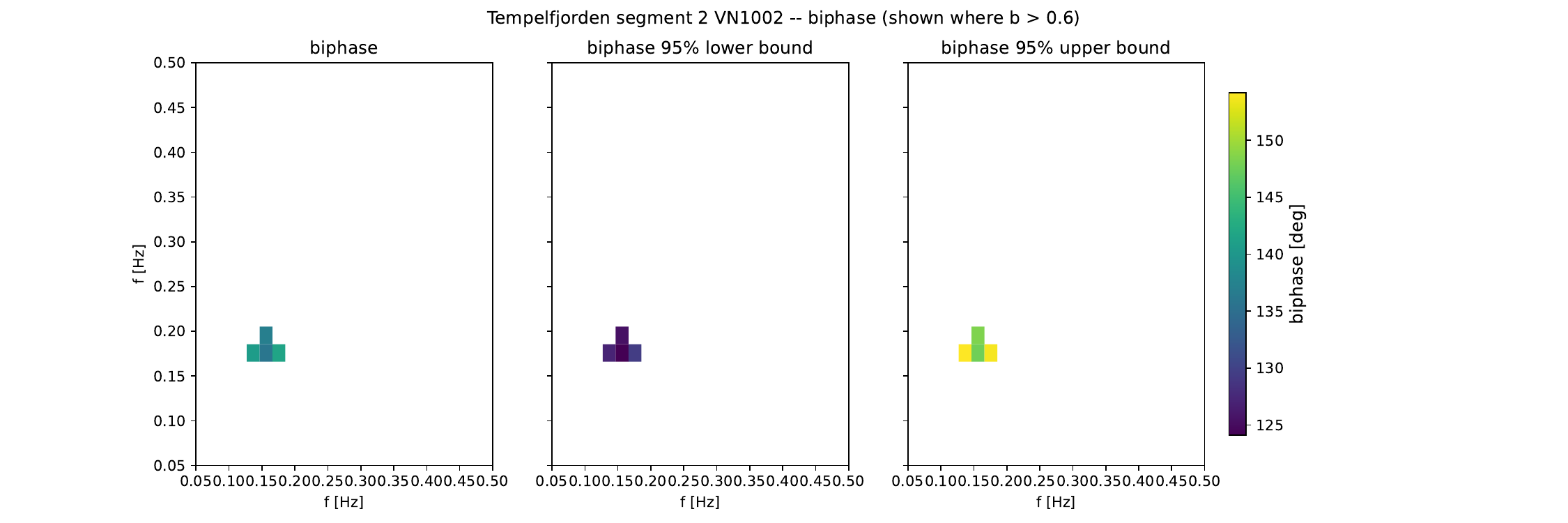}
    \includegraphics[width=0.78\linewidth]{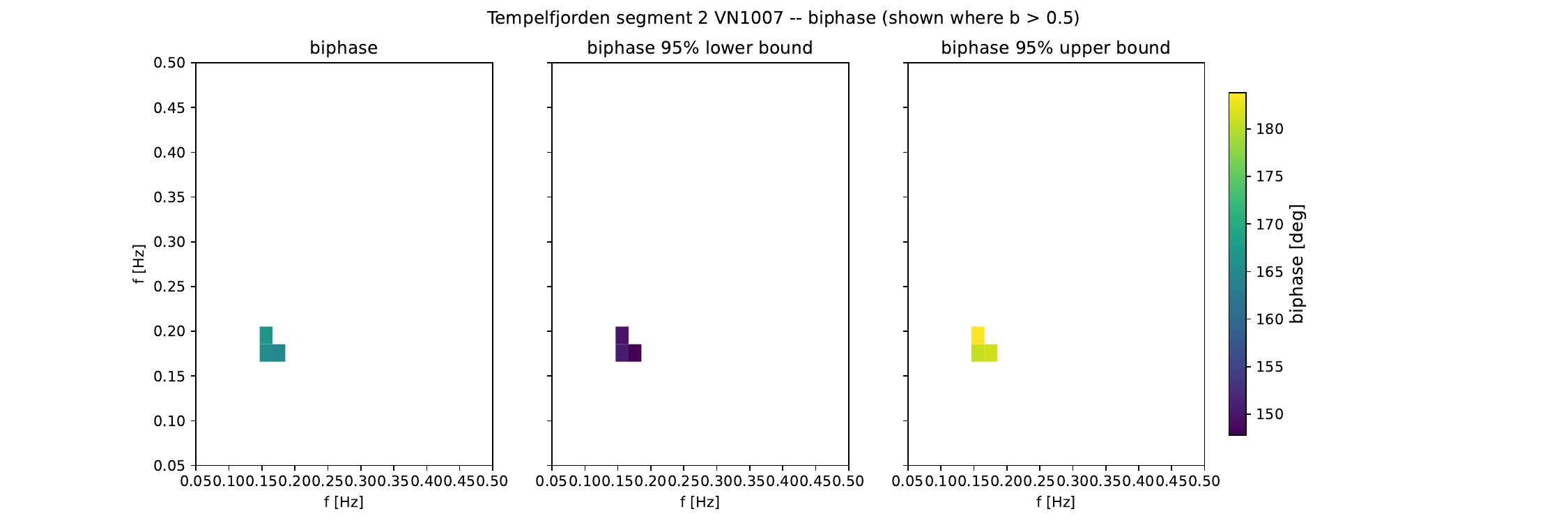}
    \includegraphics[width=0.78\linewidth]{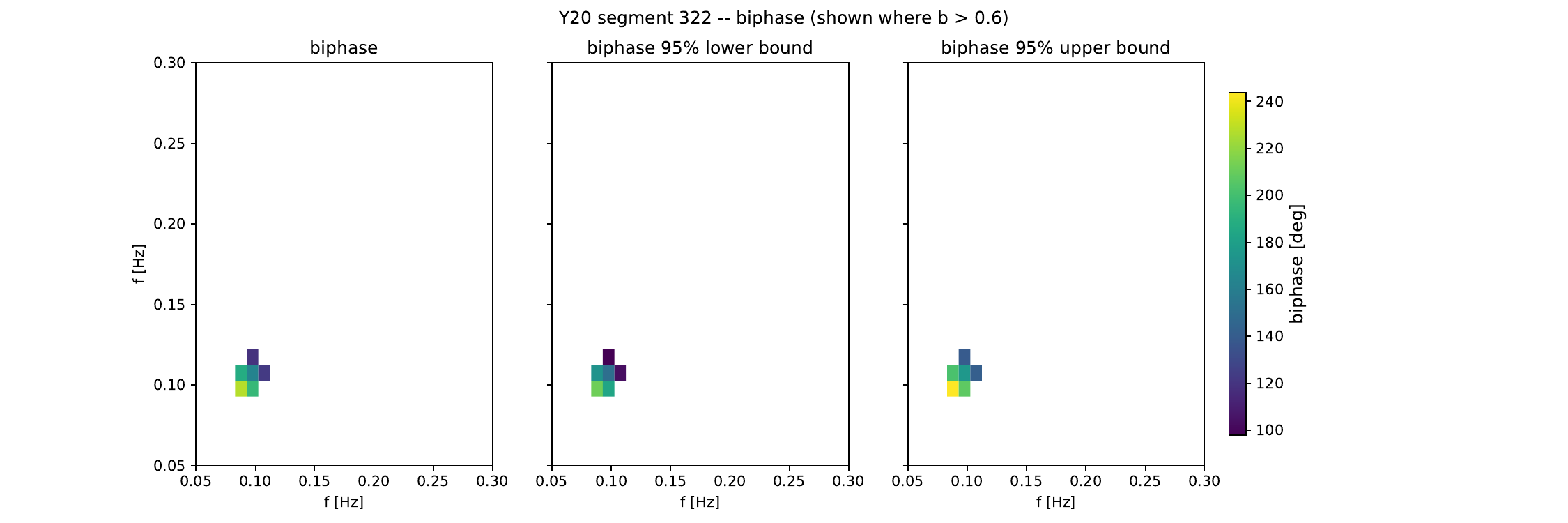}
    \caption{Biphase information, shown only for the bins where the bicoherence is high enough that it is meaningful to consider. Top to bottom show the results for VN1002 segment 2 (T15 deployment), VN1007 segment 2 (T15), and the Y20 deployment segment 322. For each row, the left column shows the biphase estimate, and the middle and right columns show the 95\% confidence interval range (under the assumption of a Gaussian stochastic process). Observe that all results are clearly different from the 0-phase lock-in that would be expected for the classical Stokes harmonic bound wave.}
    \label{fig_biphase_T15}
\end{figure}

Fig. \ref{fig_biphase_T15} shows the biphase for the T15 dataset, segment 2 (similar to what is discussed in Fig. \ref{fig:t15_bicoherence}), and for the Y20 dataset, segment 322 (similar to Fig. \ref{fig:y20_bicoherence}).

\noappendix       




\appendixfigures  

\appendixtables   


\authorcontribution{\textbf{Conceptualization}: J.R. \textbf{Data collection and curation}: J.R., J.V., T.N., G.S., T.K., A.J., Ø.B., G.H., M.M., A.M., B.W., K.H.C., P.H. \textbf{Formal analysis and Software}: J.R., J.V. \textbf{Funding acquisition}: J.R., J.V., A.B., T.W., A.J., Ø.B., M.M., K.H.C. \textbf{Investigation}: J.R., J.V., T.N., G.S., A.B., T.W., L.W.D., A.M., K.T. \textbf{Methodology}: J.R, A.B., T.W., J.V, T.N. \textbf{Project administration}: J.R., A.B., T.W., A.J., Ø.B., M.M., L.W., A.M., P.H. \textbf{Validation}: J.R., J.V., T.N. \textbf{Visualization}: J.R., J.V., T.N. \textbf{Writing - original draft preparation}: J.R., with contributions from J.V. \textbf{Writing - review \& editing}: all.} 
 
\competinginterests{Some authors are members of the editorial board of the journal "The Cryosphere".} 

\codedataavailability{The data and code used in this manuscript are made openly available, see the Appendix A for more details.}

\begin{acknowledgements}
The help of the UNIS logistics team during the 2015 T15 fieldwork is gratefully acknowledged. The help of laboratory engineer Olav Gundersen (UiO) when building and preparing the equipment used in the T15 and Y20 deployments is gratefully acknowledged. We would like to thank the crew of the RV Kronprins Haakon for invaluable help deploying the buoys in the Arctic. The authors thank Pr. Yngve Kristoffersen (UiB) for having deployed the buoy used in the Y20 dataset. Funding from the Research Council of Norway through the projects "Waves in Oil and Ice" (WOICE, project 233901), "Dynamics Of Floating Ice" (DOFI, project 280625), "Arven Etter Nansen" (AeN, project 276730), "FOCUS" (project 301450), is gratefully acknowledged. Support was also received from the Australian Antarctic Program (AAS4496, AAS4506, AAS4625) and the Australian government (ASCI000002).
\end{acknowledgements}

\bibliographystyle{copernicus}
\bibliography{references.bib}

\end{document}